\documentclass[%
 reprint,
 nofootinbib,
 aps,
 prd,
]{revtex4-2}

\usepackage{graphicx}
\usepackage[dvipsnames]{xcolor}
\usepackage{amsmath}
\usepackage{amssymb}
\usepackage{amsfonts}
\usepackage[ISO]{diffcoeff}
\usepackage{mleftright}
\usepackage{siunitx}
\usepackage[normalem]{ulem}
\usepackage{placeins}
\usepackage{booktabs}
\usepackage{csquotes}
\usepackage{hyperref}
\usepackage[capitalise]{cleveref}

\usepackage{aas_macros}

\makeatletter
\AtBeginDocument{\immediate\write\@auxout{\string\citation{apsrev42Control}}}
\makeatother

\DeclareSIUnit{\pc}{pc}
\DeclareSIUnit{\kpc}{\kilo\pc}
\DeclareSIUnit{\Mpc}{\mega\pc}
\DeclareSIUnit{\yr}{yr}
\DeclareSIUnit{\Myr}{\mega\yr}
\DeclareSIUnit{\Gyr}{\giga\yr}
\DeclareSIUnit{\Msun}{\text{\ensuremath{M_\odot}}}
\DeclareSIUnit{\hHubble}{\text{\ensuremath{h}}}
\crefname{section}{\S}{\S}
\crefformat{section}{\S#2#1#3}
\crefmultiformat{section}{\S#2#1#3}{ and~\S#2#1#3}{, \S#2#1#3}{ and~\S#2#1#3}
\AddToHook{cmd/appendix/before}{%
    \crefalias{chapter}{appendix}%
    \crefalias{section}{appendix}%
    \crefalias{subsection}{appendix}%
}

\definecolor{RedWine}{rgb}{0.743,0,0}
\definecolor{GrassGreen}{rgb}{0.125,0.75,0.125}
\definecolor{RoyalBlue}{rgb}{0.25,0.41,0.88}
\definecolor{DarkCyan}{rgb}{0,0.5,0.5}

\newcommand*{\codename}{\textsc{KiSS-SIDM}}
\newcommand*{\kns}{\ensuremath{{\mathrm{Kn}}_{\mathrm{s}}}}

\begin{document}

\preprint{APS/123-QED}

\title{Gravothermal Evolution of Self-Interacting Dark Matter Halos across Knudsen Numbers}

\author{James Gurian}
\email[\ttfamily]{jgurian@carnegiescience.edu}
\affiliation{%
Carnegie Institution for Science, 813 Santa Barbara Street, Pasadena, CA 91101, USA and\\
Perimeter Institute for Theoretical Physics, 31 Caroline Street N, Waterloo, ON N2L 2Y5, Canada%
}%
\author{Suroor Seher Gandhi}
\affiliation{%
Perimeter Institute for Theoretical Physics, 31 Caroline Street N, Waterloo, ON N2L 2Y5, Canada%
}%
\author{Neal Dalal}
\affiliation{%
Perimeter Institute for Theoretical Physics, 31 Caroline Street N, Waterloo, ON N2L 2Y5, Canada%
}%
\author{Simon May}
\affiliation{%
Fakultät für Physik, Universität Bielefeld, Universitätsstraße 25, 33615 Bielefeld, Germany%
}


\begin{abstract}
If the dark matter self-interacts elastically via a light mediator, then at relative velocities above a scale $w$ dictated by particle physics parameters, the cross section scales as $\sigma \propto v^{-4}$. Depending on the value of $w$, the self-interaction cross section in low-mass halos could be (much) larger than has been previously studied extensively. We perform kinetic simulations of isolated self-interacting dark matter halos using the code \textsc{KiSS-SIDM} for a wide range of halo masses, concentrations, and cross sections from $50 \,\mathrm{cm}^2 \, \mathrm{g}^{-1}$ to $250{,}000 \, \mathrm{cm}^2 \, \mathrm{g}^{-1}$, including several velocity-dependent and fully differential (anisotropic) cross sections. We demonstrate that for idealized NFW halos the collapse time can be predicted from the Knudsen number at the scale radius of the halo. We perform similar calculations for halos extracted from cosmological simulations using either $N$-body (collisionless) or SPH (strongly collisional) dynamics, and then evolve cutouts with \textsc{KiSS-SIDM}. For these halos, the collapse times differ from the idealized NFW simulations by up to a factor of a few. Further, we calibrate a hydrostatic fluid model to these results, which solves the structure equations as a boundary value problem via two-point shooting. The fluid calculation predicts that for velocity-independent cross sections, a convective region of the halo develops late in the collapse. The collapse times agree well between the kinetic simulations and the fluid code, especially in the velocity-independent case. The density profiles agree well in some cases, which include all of the velocity-dependent cross section runs.
\end{abstract}


\maketitle


\section{Introduction}
\label{sec:intro}

The standard cold dark matter (CDM) model assumes that the dark matter is effectively collisionless. This model is an idealization, and more detailed theories that explain the abundance of dark matter generally involve non-minimal interactions beyond gravity.  These interactions can take many possible forms, such as a quartic $\lambda\phi^4$ self-interaction for scalar dark matter \cite[e.\,g.][]{Peebles_Vilenkin_1999, Goodman_2000, Peebles_2000, Bento_2000}, or a Yukawa-like $g \phi \bar{\chi} \chi$ interaction between DM and a scalar mediator \cite[e.\,g.][]{Loeb_2011, Arkani_Hamed_2009}. The latter interaction can give rise to a self-interaction between $\chi$ particles mediated by $\phi$ particles that conserves number, i.\,e.\ elastic scattering, commonly referred to as self-interacting dark matter (SIDM) \cite{Buckley_2010, Feng_2010, Tulin_2013}. In the Born regime, this Yukawa-like elastic scattering can be described by a velocity-dependent and angle-dependent cross section \cite{Yang_2022, Outmezguine:2022bhq}
\begin{equation}
    \diff{\sigma}{\Omega} = C 
    \left[
    \left(\frac v c \sin\mleft(\frac{\theta}{2}\mright) \right)^2 
    + \left(\frac{m_\phi}{m_\chi}\right)^2
    \right]^{-2},
\end{equation}
where $m_\phi$ and $m_\chi$ are the masses of the mediator and DM particle, $v$ is the relative velocity between the pair of DM particles, and $C$ is the normalization constant related to $g$ and the other parameters.
Analysis of galaxy clusters strongly constrains the self-interaction cross section at the relative velocity scales typical to clusters, $v_{\mathrm{rel}} \approx \qty{1000}{\km\per\s}$, to be $\sigma/m_\chi < \qty{1}{\cm\squared\per\g}$ \cite{markevitch_direct_2004, Andrade_2021, Robertson_2016}. 
For heavy mediators $m_\phi \gg m_\chi$, this interaction behaves like hard-sphere scattering where the cross section is isotropic and independent of velocity, and in this regime the above cluster bounds would also apply to all smaller objects like galaxies.  
However, if the self-interaction is mediated by a light particle with $m_\phi \ll m_\chi$, the cross section would be strongly velocity-dependent, $\sigma \propto v^{-4}$ at velocities where $v/c \gg m_\phi/m_\chi$. In this scenario, a cross section with interesting phenomenological consequences remains viable at the scale of dwarf galaxies, which can have $v_{\mathrm{rel}} < \qty{100}{\km\per\s}$ \cite{Tulin_2018, adhikari2022astrophysical}.

In these environments, a cross section of \qtyrange{10}{100}{\cm\squared\per\g} can thermalize the inner regions of initially Navarro-Frenk-White (NFW) \cite{1997ApJ...490..493N} dwarf galaxy dark matter halos in less than a Hubble time, which leads to the formation of a nearly constant-density core \cite[e.\,g.][]{Vogelsberger_2012, Vogelsberger_2014, Fry_2015, Elbert_2015}. This is to be contrasted with the \enquote{cuspy} \cite{Bullock_2017} NFW profile:
\begin{equation}
    \rho(r) = \frac{\rho_{\mathrm{s}}}{\frac{r}{r_{\mathrm{s}}} \mleft(1 + \frac{r}{r_{\mathrm{s}}}\mright)^2}.
\end{equation}
Sustained conduction of heat by self-scatterings from the isothermal core to the cool envelope leads to a runaway collapse. Absent any non-thermal sources of pressure, the collapse is expected to continue until a black hole forms. This so-called \enquote{gravothermal collapse} has been the subject of significant work \cite{Balberg_2002, Gad-Nasr:2023gvf, Gu:2026zzq, zeng2026feedingvoidcoevolutionsidmseeded, Meng:2026hzv}, but its ultimate stable endpoint largely remains an open question. 

Due to this phenomenology, some research suggests that a non-zero self-scattering cross section is preferred over CDM at galaxy scales \cite{Kamada2017, Nadler_2023, yang_2023a, Grant25, Yu_2026}.
In this literature, it is commonly assumed that $m_\phi c/m_\chi \approx \qty{100}{\km\per\s}$, so that the SIDM cross section attains its maximum value of $\lesssim \qty{100}{\cm\squared\per\g}$ around the velocity scale of dwarf galaxies. From one perspective, this scenario may seem finely tuned: there is no obvious connection between the masses of the dark matter particle and mediator and the velocity dispersion scale in dwarf galaxies. Specifically, this possibility already requires a large hierarchy between the scalar mediator mass and the DM mass, with $m_\phi/m_\chi \approx 10^{-4}$, and there is no obvious reason why this value for the ratio of masses would be preferred over some other value like $10^{-5}$ or $10^{-6}$. 
Without this fine tuning, at velocity scales somewhat smaller than those of classical dwarfs (e.\,g., at $\approx \qty{10}{\km\per\s}$ in ultrafaint dwarfs), the SIDM cross section could be \emph{much} larger than that invoked to explain the observed rotation curves of classical dwarfs. 

In either scenario, a prediction of SIDM at these cross sections is that some halos will undergo runaway core collapse. Extensive recent work has focused on the numerical difficulties in accurately simulating this phenomenology, with traditional $N$-body methods demonstrating poor energy conservation and non-converged collapse timescales \cite{palubski2024numericalchallengesmodelinggravothermal, mace2024convergencetestsselfinteractingdark, fischer2025accuratelysimulatingcorecollapseselfinteracting}. \citet{Gurian_2025} introduced a new simulation code, {\codename} (Kinetic Spherically Symmetric SIDM), which can address the aforementioned computational issues. {\codename} is based on the Direct Simulation Monte Carlo (DSMC) framework from computational fluid dynamics, and integrates the orbits of dark matter particles subject to stochastic, pairwise scattering. It assumes spherical symmetry, but resolves the full 3D velocity distribution (see also \cite{kamionkowski2025numericalevolutionselfgravitatinghalos}). Compared to other $N$-body codes, it is less computationally expensive and more accurate in the core collapse regime. 

Meanwhile, a related literature has focused on studying the late stages of the gravothermal collapse by adopting a fluid approximation. This approach (formally identical to the problem of hydrostatic stellar structure) involves solving the continuity, hydrostatic equilibrium, luminosity, and energy transport equations for a sequence of quasi-static profiles. The key assumption is that a suitable ansatz exists for the conductive luminosity, which closes the moment hierarchy of the collisional Boltzmann equation. This body of work has focused on identifying universal behaviors across different cross sections and halos which can be used to predict the collapse time and the properties of the core deep into the collapse \cite{Balberg_2002a, Pollack2015, Essig_2019, Nishikawa_2020, Outmezguine:2022bhq, Gad-Nasr:2023gvf}.

The \enquote{universal} evolution suggested by these studies has not been validated at higher cross sections. In particular, an oft-referenced inverse relationship between the collapse time and the cross section ($t_{\mathrm{col}} \propto 1/\sigma$) is known to break down at high cross sections. This is because the collapse time $t_{\mathrm{col}}$ decreases with increased thermal conductivity $\kappa$, and $\kappa$ varies non-monotonically with the cross section $\sigma$. The relevant parameter which determines the cross section dependence of the conductivity is the Knudsen number. This is the ratio of the particle mean free path (MFP) $\lambda$ to the scale of spatial variation in the problem (the gravitational scale height $H$):
\begin{equation}
    \mathrm{Kn} \equiv\frac{\lambda}{H} =\frac{1}{\rho\sigma_{\mathrm{m}}}\sqrt{\frac{4 \pi G \rho}{\nu^2}},
\end{equation}
where $\rho$ is the dark matter mass density, $\sigma_{\mathrm{m}}$ is the (local) value of the cross section per unit mass, $G$ is the gravitational constant, and $\nu$ is the 1D velocity dispersion. At large $\mathrm{Kn}$ (the long mean free path, or LMFP regime), scattering is infrequent but efficiently transports energy across the system and the thermal conductivity is proportional to the cross section, $\kappa \propto\sigma_{\mathrm{m}}$. Since the collapse is sped up by efficient heat transport, the aforementioned relationship $t_{\mathrm{col}}\propto 1/\sigma_{\mathrm{m}}$ holds. However, at small Kn (the short mean free path, SMFP, regime), scattering is frequent but unable to efficiently transport energy due to the short mean free path and $\kappa \propto 1/\sigma_{\mathrm{m}}$. This slows down the collapse, making $t_{\mathrm{col}} \propto \sigma_{\mathrm{m}}$. Thus, for any given halo there exists an intermediate cross section---or an intermediate $\mathrm{Kn}$---that maximizes the thermal conductivity $\kappa$ and thereby minimizes the gravothermal collapse timescale $t_{\mathrm{col}}$. 

In this work, we perform simulations with {\codename} of a wide range of halo masses, concentrations, and cross sections. We show that for idealized, isolated NFW initial halo profiles, the gravothermal collapse timescale follows a universal trend as a function of Kn with relatively little scatter. This universality holds across the SMFP to LMFP transition, and deep into each regime, independent of the halo concentration. We find that the collapse time is well-fit by a sum of SMFP and LMFP terms, with a log-normal bump in the intermediate regime, \cref{eq:tcolfit}. 

However, the assumption of an NFW initial profile is not justified in the SMFP regime, because frequent scattering will alter the hierarchical assembly of the halo. Further, cosmological accretion and mergers are known to modify the collapse timescale. As a preliminary exploration of the effects of strongly collisional SIDM on structure formation, we perform matched DM-only $N$-body (collisionless) and smoothed particle hydrodynamics simulations (where the dark matter behaves as an ideal fluid with no energy transport). We then simulate the collapse phase using {\codename} for a halo extracted from these cosmological volumes. 
We find deviations of up to a factor of a few from the trend in $t_{\mathrm{col}}$ as a function of $\mathrm{Kn}$ identified for isolated, idealized NFW halos.

Finally, we present a new conducting fluid code, \textsc{SIDMStars}, for an improved treatment of the core collapse phase within the fluid approximation. In the SIDM literature, a standard numerical approach is to solve the structure equations for a hydrostatic halo by a finite difference update of the internal energy in discrete mass shells, followed by iteratively solving the linearized equation of hydrostatic equilibrium (at fixed specific entropy) \cite{Balberg_2002, Pollack2015, Nishikawa_2020, Outmezguine:2022bhq, Gad-Nasr:2023gvf, mace2025calibratingsidmgravothermalcatastrophe}. We instead solve for the halo structure at each time step as a boundary value problem via shooting. By comparing the results of this code with those from the {\codename} simulations, we calibrate the interpolation between the conductivities in the LMFP and SMFP regimes. With this calibration, the collapse time agrees reasonably well between the kinetic ({\codename}) and fluid (\textsc{SIDMStars}) results across $\mathrm{Kn}$.

In this work, we simulate a very wide range of cross sections. We emphasize that although we simulate, for example, a cross section of \qty{e4}{\cm\squared\per\g} in a \qty{e9}{\Msun} halo, we do not make any specific claim that this is consistent with observations. The aim of this work is simply to understand the gravothermal dynamics of dark matter halos across collisional regimes. Doing so may have implications for hierarchical structure formation in dark matter models where the dark matter is strongly collisional at small scales and low velocity dispersions (which are not currently subject to direct observational constraints). This includes not only the light mediator-type models considered here but also other dark sectors. For comparison, the \enquote{self-interaction} cross section of ordinary hydrogen in its bound state is $\approx \qty{e7}{\cm\squared\per\g}$. Further, even though some of the parameter space studied here is ruled out by observations, correctly understanding the phenomenology can help correctly delineate these constraints. We apply the results of this paper to constrain the allowed parameters in the light mediator SIDM model in a companion paper.

This paper is organized as follows. In \cref{sec:kin_sim} we describe the suite of idealized (NFW initial profile) collapse simulations we have conducted using {\codename}, and present fitting formulae to the collapse time as a function of $\mathrm{Kn}$. In \cref{sec:comsosim}, we explore the impact of strong self-interactions on structure formation by dropping the assumption of an initial NFW profile and conduct matched simulations of a cosmological box in the collisionless and fluid limits. We then simulate the core collapse in a cutout halo using {\codename}, extracted from the cosmological simulations. In \cref{sec:fluid}, we introduce a new conducting fluid code, \textsc{SIDMStars}, which solves the halo structure equations via shooting, and employs a thermal conductivity whose form is calibrated to the kinetic simulations in \cref{sec:kin_sim}. We conclude and discuss future and related work in \cref{sec:end}.

\section{Kinetic Simulations of Idealized Halos}
\label{sec:kin_sim}

\subsection{Numerical Setup}

We use {\codename} to simulate the gravothermal evolution of isolated, idealized dark matter halos across a range of halo masses, concentrations, and cross sections. This suite of simulations includes various angle- and velocity-dependencies as well (see \cref{tab:halo_parameters} for a full list). $\codename$  is a spherically symmetric, kinetic code developed specifically for studying core collapse in SIDM halos, based on the Direct Simulation Monte Carlo (DSMC) approach \cite{Gurian_2025}. Assuming spherical symmetry greatly reduces the computational cost and improves the energy accuracy compared to traditional $N$-body methods, enabling simulation deep into the core-collapsed regime even at large ($\gg \qty{100}{\cm\squared\per\g}$) self-interaction cross sections. 
We have extended {\codename} to handle velocity-dependent and anisotropic scattering, elaborated in \cref{app:vdep_imp}. 

In all cases considered in this section, the initial conditions are idealized in the sense that they are for an isolated, truncated NFW profile \cite{NFW} generated using \textsc{SpherIC} \cite{Garrison_Kimmel_2013}. Most runs use one million particles, with a few simulations at higher and lower particle numbers. While many of our simulations employ velocity-independent, isotropic cross sections, we carry out several runs with velocity-dependent isotropic cross sections, as well as with fully differential cross sections.  

For the velocity-dependent isotropic runs, we use the cross section per unit mass:
\begin{equation}
    \sigma_{\mathrm{m}}(v) = \frac{\sigma_0}{(1+(v/w)^2)^2},
    \label{eq:rv4}
\end{equation}
with $\sigma_0$ the cross section per unit mass at low relative velocity and  $w = \qty{1}{\km\per\s}$. For the fully differential runs (i.\,e., $\sigma_{\mathrm{m}} = \sigma_{\mathrm{m}}(v,\theta)$), we use the package \href{https://github.com/dangilman/pykawa}{\texttt{pykawa}}\footnote{\url{https://github.com/dangilman/pykawa}} to generate the differential cross section for an attractive Yukawa interaction with $m_\phi = \qty{416}{\keV}$, $m_\chi = \qty{35}{\GeV}$ and $\alpha = 10^{-2.7}$, which is the default in \texttt{pykawa}.

\begin{table*}
    \centering
    \caption{Summary of halo properties, scattering cross sections, and particle numbers of the kinetic (\codename) simulations.
    $M_{200\mathrm{c}}$ encloses a mean density of $200\rho_{\mathrm{crit}}(z)$ in the initial conditions (Planck~2018 cosmology), at the redshift $z$ listed; $c = r_{200\mathrm{c}}/r_{\mathrm{s}}$. $M_{\mathrm{tot}}$ is the total mass of all particles simulated, and $\tilde{\sigma}_{m}$ is the viscosity-weighted effective cross section (\cref{eq:eff}) evaluated at $\nu_{\mathrm{eff}} = 0.71\,v_{\max}$. Regularized $v^{-4}$ runs use the cross section of \cref{eq:rv4} with $w = \qty{1}{\km\per\s}$.}
    \smallskip
    \setlength{\tabcolsep}{5pt}
    \begin{tabular}{l c *{2}{S[table-format=1.2e1]} S[table-format=2.2] S[table-format=1.2e+1] S[table-format=4.0] S[table-format=1.2e1] S[table-format=1.2e+1] l S[table-format=1.1e1]}
    \toprule
    Halo & $z$ & {$M_{200\mathrm{c}}$} & {$M_{\mathrm{tot}}$} & {$c$} & {$\rho_{\mathrm{s}}$} & {$r_{\mathrm{s}}$} & {$\tilde{\sigma}_{\mathrm{m}}$} & {$\kns$} & Type & {$N$} \\
     & & {(\unit{\Msun})} & {(\unit{\Msun})} & & {(\unit{\Msun\per\pc\cubed})} & {(\unit{\pc})} & {(\unit{\cm\squared\per\g})} & & & \\
    \midrule
    \textbf{Halo 1} & 4 & 3.25e6 & 1.95e7 & 2.98 & 1.40e-2 & 308
        & 1.67e2  & 6.65e0   & regularized $v^{-4}$  & e6 \\
    & & & & & & & 2.50e2  & 4.44e0   & constant              & e6 \\
    & & & & & & & 5.03e2  & 2.21e0   & attractive Yukawa     & e6 \\
    & & & & & & & 5.38e2  & 2.07e0   & constant              & e6 \\
    & & & & & & & 1.00e3  & 1.11e0   & constant              & e6 \\
    & & & & & & & 5.00e3  & 2.22e-1  & regularized $v^{-4}$  & e6 \\
    & & & & & & & 7.48e3  & 1.48e-1  & constant              & e6 \\
    & & & & & & & 1.24e4  & 8.98e-2  & regularized $v^{-4}$  & e6 \\
    & & & & & & & 2.08e4  & 5.34e-2  & regularized $v^{-4}$  & e6 \\
    & & & & & & & 2.50e4  & 4.44e-2  & constant              & e6 \\
    & & & & & & & 5.57e4  & 2.00e-2  & regularized $v^{-4}$  & e6 \\
    & & & & & & & 8.33e4  & 1.33e-2  & constant              & e6 \\
    & & & & & & & 1.67e5  & 6.65e-3  & regularized $v^{-4}$  & e6 \\
    & & & & & & & 1.67e5  & 6.65e-3  & regularized $v^{-4}$  & 3e6 \\
    & & & & & & & 2.50e5  & 4.44e-3  & constant              & e6 \\

    \addlinespace

    \textbf{Halo 2} & 0 & 1.15e9 & 1.50e9 & 18.7 & 2.73e-2 & 1180
        & 5.00e1  & 2.97e0   & constant             & e6 \\
    & & & & & & & 3.86e2  & 3.85e-1  & regularized $v^{-4}$ & e6 \\
    & & & & & & & 5.00e2  & 2.97e-1  & constant             & e6 \\
    & & & & & & & 1.00e3  & 1.48e-1  & constant             & e6 \\
    & & & & & & & 1.93e3  & 7.69e-2  & regularized $v^{-4}$ & e6 \\
    & & & & & & & 1.00e4  & 1.48e-2  & constant             & e6 \\

    \addlinespace

    \textbf{Halo 3} & 4 & 1.61e8 & 3.10e8 & 5.86 & 6.28e-2 & 576
        & 7.32e1  & 1.81e0   & attractive Yukawa     & e6 \\
    & & & & & & & 3.54e2  & 3.74e-1  & regularized $v^{-4}$ & e6 \\
    & & & & & & & 2.14e3  & 6.20e-2  & constant             & e6 \\
    & & & & & & & 1.97e4  & 6.72e-3  & regularized $v^{-4}$ & 2.5e5 \\

    \addlinespace

    \textbf{Halo 4} & 0 & 3.06e7 & 3.24e7 & 26.4 & 6.83e-2 & 250
        & 8.00e1  & 3.51e0   & constant             & 2e6 \\
    & & & & & & & 5.00e2  & 5.61e-1  & constant             & 2e6 \\
    
    \bottomrule
    \end{tabular}
    \label{tab:halo_parameters}
\end{table*}

\subsection{Timescales and Qualitative Evolution}

Energy transport in SIDM halos is characterized by the thermal conductivity, which relates the heat flux to the temperature gradient. In this work we use velocity-squared units for the temperature throughout,\footnote{Via the velocity $v$ associated with any temperature through $T = m v^2/k_{\mathrm{B}}$ for a fixed particle mass $m$. In other words, we implicitly express any temperature $T$ using the equivalent quantity $\tilde{T} = k_{\mathrm{B}}T/m$.} and define the conductivity $\kappa$ via
\begin{equation}
\label{eq:q}
    q = -\kappa \diff{u}{r},
\end{equation}
where $q$ is the heat flux (energy/area/time), and $u = (3/2)\nu^2$ is the internal energy with the 1D velocity dispersion $\nu$ (assuming a monatomic ideal gas equation of state). The conductivity is a function of the cross section, the local velocity dispersion, and (except in the deep short mean free path regime) the local density.

Large ($\gtrsim \qty{1}{\cm\squared\per\g}$) isotropic, velocity-independent cross sections are restricted by strong constraints on the effective cross section at cluster scales \cite{markevitch_direct_2004, Andrade_2021}. However, \citet{Yang_2022} demonstrated that for halos in which the mean free path is initially large compared to the gravitational scale height (i.\,e., halos that start out in the LMFP regime), the effects of velocity- and angle-dependent scattering on the collapse time can be absorbed by an appropriately moment-weighted, isotropic cross section per unit mass (Eq.~(4.2) in \cite{Yang_2022}):
\begin{widetext}
\begin{equation}
    \label{eq:eff}
    \tilde{\sigma}_{\mathrm{m}} = \frac{1}{512 \nu_{\mathrm{eff}}^8 } \int v^2 \dl v \dl\cos(\theta) \diff{\sigma_{\mathrm{m}}}{\cos(\theta)} v^5 \sin(\theta)^2 \exp\mleft(-\frac{v^2}{4 \nu_{\mathrm{eff}}^2 }\mright),
\end{equation}
\end{widetext}
where $\nu_{\mathrm{eff}}$ is an effective one-dimensional velocity dispersion and $\difs{\sigma_{\mathrm{m}}}{\cos(\theta)}$ is the differential cross section per unit mass. Usually $\nu_{\mathrm{eff}} = 0.64 \, v_{\mathrm{max}}$ with $v_{\mathrm{max}}$ the maximum circular velocity in the halo \cite{Yang_2022, Outmezguine:2022bhq}, which corresponds to the velocity dispersion at maximum core expansion. We find that our velocity-dependent simulations align better with the velocity-independent simulations under $\nu_{\mathrm{eff}}=0.71v_{\mathrm{max}}$, and adopt this throughout. (This is also the reason that most of the velocity-dependent isotropic cross sections are not round numbers: the simulations targeted a given $\tilde{\sigma}_{\mathrm{m}} $ under the $0.64 v_{\mathrm{max}}$ definition and were relabeled to the $0.71 v_{\mathrm{max}}$ values.)

Although we only simulate four halos, these results have a more general applicability, because the Knudsen number is the true controlling parameter in the problem. That is, the problem is unchanged after a rescaling to the same Knudsen number at different halo mass and cross section, a fact which will be demonstrated shortly (\cref{fig:sim_data_const,fig:tcol_tdyn_all}). 

A further motivation for studying such large cross sections is that the thermal conductivity in the SMFP regime can be derived in the Chapman–Enskog expansion \cite{Chapman70}. Specifically, the conductivity $\kappa_{\mathrm{SMFP}}$ can be expressed in terms of thermal moments of the cross section:
\begin{equation}
    \label{eq:Gammap}
    \Gamma_{p}(\nu) \equiv \frac{\langle v^p \sigma_{\mathrm{v,m}}(v) \rangle}{\langle v^p\rangle},
\end{equation}
where $p$ is an integer, the angle brackets $\langle\cdot\rangle$ denote the thermal average over the relative velocity distribution at 1D velocity dispersion $\nu$, and 
\begin{equation}
    \label{eq:sigmavm}
    \sigma_{\mathrm{v,m}} =\frac{3}{2} \int \dl\Omega \sin(\theta)^2 \diff{\sigma_{\mathrm{m}}}{\Omega}   
\end{equation}
is the viscosity cross section per unit mass.  Note that \cref{eq:eff} can be expressed as $\tilde{\sigma}_{\mathrm{m}} = \Gamma_5(0.71 v_{\mathrm{max}})$. In terms of $\Gamma_p$, the SMFP conductivity is
\begin{equation} 
    \label{eq:kappa_smfp}
    \kappa_{\mathrm{SMFP}} = \frac{b\,\nu}{\Gamma_{\mathrm{eff}}},
\end{equation}
with $b=25 \sqrt{\pi}/{32}$, and $\Gamma_{\mathrm{eff}}$ is given in \cite{Gad-Nasr:2023gvf} (at second order in the Chapman–Enskog expansion) as:%
\footnote{The dimensionful $\Gamma_{\mathrm{eff}}$ differs from the dimensionless $K_{\mathrm{eff}}$ of \cite{Gad-Nasr:2023gvf, Outmezguine:2022bhq} by a factor of $\sigma_{\mathrm{v,m}}$ (\cref{eq:sigmavm}), the value of the cross section at some velocity scale.}
\begin{equation}
   \Gamma_{\mathrm{eff}} \approx\ \Gamma_{\mathrm{eff}}^{(2)} \equiv \frac{28 \Gamma_5^2 + 80 \Gamma_5\Gamma_9 -64 \Gamma_7^2}{77 \Gamma_5 -112 \Gamma_7 +80 \Gamma_9}.
    \label{eq:g2}
\end{equation}

Meanwhile, in the LMFP regime, the conductivity is \cite{Lynden-Bell80, Outmezguine:2022bhq}:
\begin{equation}
    \label{eq:kappa_lmfp}
    \kappa_{\mathrm{LMFP}} = \frac{a C}{4 \pi G}\Gamma_5 \rho \nu^3,
\end{equation}
with $a = 4/\sqrt{\pi}$ and $C$ a constant which must be calibrated via simulations.

Finally, in the intermediate mean free path regime there is no such controlled expansion about a quasi-Maxwellian distribution, and the thermal conductivity can only be determined empirically. Pushing our simulations into the deep SMFP regime provides a longer \enquote{handle} on the behavior in the intermediate regime, so that legitimate intermediate mean free path behavior can be disentangled from the calibration constant $C$.

\begin{figure}
    \centering
    \includegraphics[width=\linewidth]{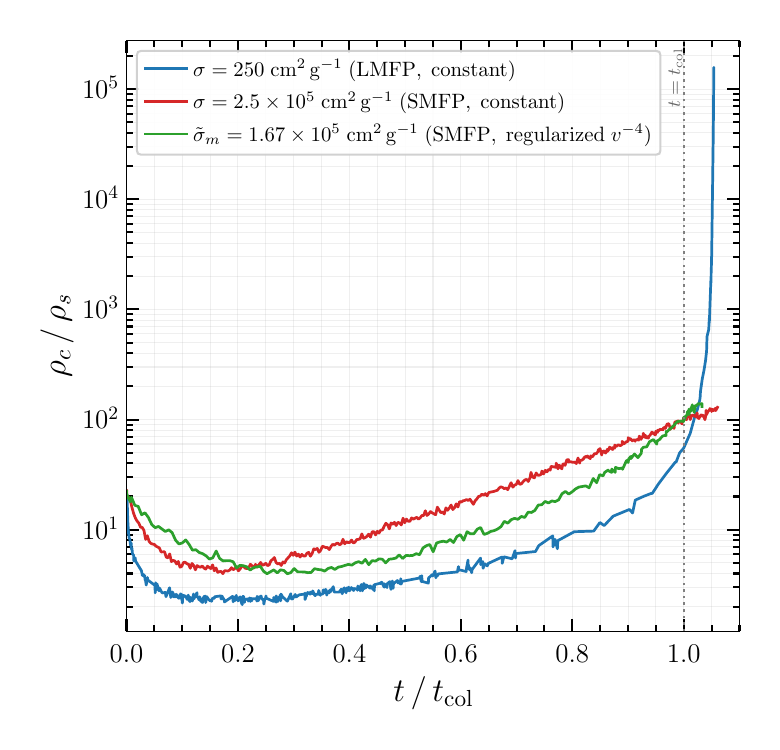}
    \caption{The evolution of the central density $\rho_{\mathrm{c}}$ for deepest LMFP (smallest $\sigma$; blue curve) and deepest SMFP (largest $\sigma$; red curve) Halo~1 constant cross section runs, along with a velocity-dependent deep SMFP (green curve) run, normalized by the scale density $\rho_{\mathrm{s}}$. Time on the horizontal axis is normalized by $t_{\mathrm{col}}$---the time required for the density to exceed 25 times its minimum value (a choice discussed in \cref{sec:collapse time}). Note the finite-time divergence of the LMFP central density, compared with the quasi-exponential growth of the SMFP density. See text below \cref{eq:t_kappa} for more discussion.}
    \label{fig:col_traj}
\end{figure}

The energy transport (Kelvin--Helmholtz) timescale is
\begin{equation}
    \label{eq:t_kappa}
    t_{\kappa} \equiv \frac{E_{\mathrm{therm}}}{L_\kappa} \approx \frac{r^2 \rho}{\kappa} \approx \frac{\nu^2}{G \kappa}
\end{equation}
where $E_{\mathrm{therm}}$ is the thermal energy within a sphere of radius $r$, and $L_\kappa$ is the conductive luminosity (the rate at which $E_{\mathrm{therm}}$ is carried out beyond $r$ by dark matter particles). In the final step, we have inserted the gravitational scale height for $r$. 
Since $\nu$ is roughly constant with $\rho$ during the collapse, in the SMFP regime, $t_\kappa$ is independent of $\rho$ (inserting \cref{eq:kappa_smfp} in \cref{eq:t_kappa}): roughly equal time per $e$-fold of density. Meanwhile, in the LMFP regime, $t_\kappa \propto 1/\rho$ (\cref{eq:kappa_lmfp} in \cref{eq:t_kappa}), so most of the time evolution is spent in the first few $e$-folds of density growth, and the runaway collapse is late and sudden. 
These differing collapse trajectories are shown in \cref{fig:col_traj} for the highest and lowest $\mathrm{Kn}$ (i.\,e., the smallest and largest cross section) simulations. Indeed, we see that the two SMFP densities (red and green curves) show slower, quasi-exponential growth with approximately equal time per $e$-fold of density growth, while the LMFP density (blue curve) diverges super-exponentially.

In practice, $\nu$ does grow slowly as $\rho$ increases over the course of the collapse, which in turn induces a $\rho$-dependence in $\kappa_{\mathrm{SMFP}}$. Thus, halos that start out in the SMFP regime can also eventually exhibit a super-exponential density runaway as the conductivity increases with increasing temperature. This is what drives the regularized $v^{-4}$ SMFP density evolution (green curve) in \cref{fig:col_traj} to deviate from the velocity-independent SMFP run (red) towards LMFP-like behavior (blue). However, the conventional wisdom that the time for the central density to run away (i.\,e.~the time to form a black hole) is dominated by the first few $e$-folds of density increase turns out to be a parameter- and model-dependent statement.

\subsection{The Collapse Time across Kn}
\label{sec:collapse time}

In the LMFP regime, the energy transport per scattering saturates, and the collapse timescale (i.\,e., the time to closely approach the density runaway or black hole formation epoch) is a fixed multiple of the relaxation (scattering) timescale, 
\begin{equation}
    t_{\mathrm{rel}}
    = \frac{\lambda_{\mathrm{s}}}{v_{\mathrm{rel}}}
    = \frac{1}{r_{\mathrm{s}} \rho_{\mathrm{s}} \tilde{\sigma}_{\mathrm{m}}}\frac{1}{\sqrt{4\pi G \rho_{\mathrm{s}}}},
\end{equation}
with the $r_s$ the NFW scale radius, such that the collapse timescale is \cite{Essig_2019}
\begin{equation}
    t_{\mathrm{col, LMFP}} \approx 200\, t_{\mathrm{rel}}\ \propto\ 1/\tilde{\sigma}_{\mathrm{m}}.
    \label{eq:tlmfp}
\end{equation}

Whereas in the SMFP regime, the conductivity is \textit{inverse} in the cross section (\cref{eq:kappa_smfp}), so that the expected scaling is,
\begin{equation}
    t_{\mathrm{col, SMFP}} \propto \tilde{\sigma}_{\mathrm{m}} \propto \tilde{\sigma}_{\mathrm{m}}^2 t_{\mathrm{rel}}.
\end{equation}
Notably, the \enquote{universal} evolution presented in previous works \cite{Essig_2019, Outmezguine:2022bhq} in the form of \cref{eq:tlmfp} holds only in the LMFP regime; it cannot hold for arbitrarily large cross sections.
A more general fit for $t_{\mathrm{col}}$ across many orders of magnitude in $\tilde{\sigma}_{\mathrm{m}}$ must be a function of the Knudsen number (say, at the scale radius), 
\begin{equation}
    \kns \equiv \frac{\lambda_{\mathrm{s}}}{r_{\mathrm{s}}}
    = \frac{1}{r_{\mathrm{s}} \rho_{\mathrm{s}} \tilde{\sigma}_{\mathrm{m}}},
\end{equation}
and should interpolate between these asymptotic behaviors. 
%

Since the density runaway is more gradual in the SMFP regime (\cref{fig:col_traj}), the collapse timescale depends more sensitively on the definition of \enquote{collapsed}.
In an initially NFW halo, the gravothermal evolution first drives heat inwards to form an isothermal core. As the isothermal region expands, the density in the core decreases to some minimum. After this point, heat travels outwards and the central density begins its runaway increase. There is not a universally agreed-upon definition of the collapse time in the literature. We take $t_{\mathrm{col}}$ as the time when the central density $\rho_{\mathrm{c}}$ exceeds its minimum value by a factor of 25: $\rho_{\mathrm{c}}/\rho_{\mathrm{c,min}}=25$. This is an imperfect definition, whose primary virtues are that this density contrast is accessible with reasonable computational cost using {\codename} for all the cases we have simulated, and that choosing some higher density threshold in the LMFP regime would shift the collapse times by $\lesssim 10\%$. In the SMFP regime, however, the collapse time defined in a different way could vary by an order-unity factor for velocity-dependent interactions, or by more for velocity-independent interactions. 

\begin{figure*}
    \centering
    \includegraphics[width=0.95\linewidth]{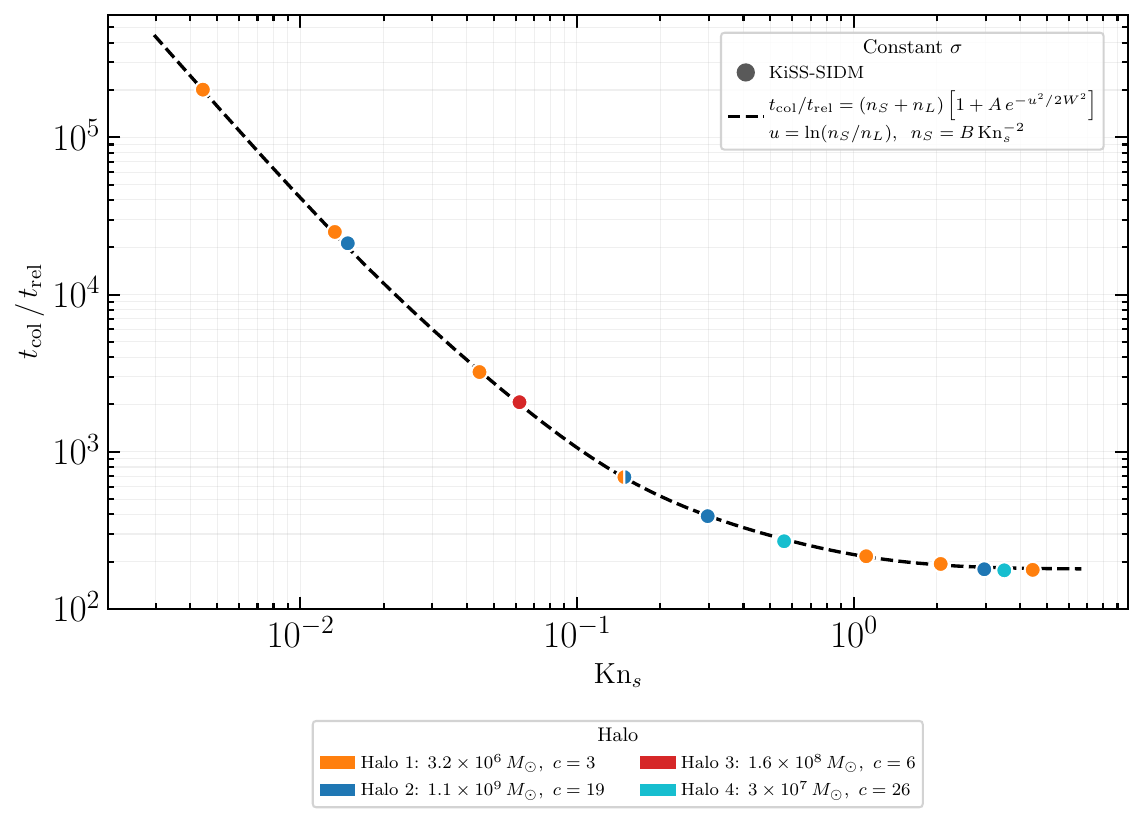}
    \caption{The collapse time (to reach $\rho_{\mathrm{c}} = 25 \times \rho_{\mathrm{c,min}}$)
    for the isotropic, velocity-independent kinetic simulations (points) and the model fit (dashed, black). As discussed in \cref{sec:collapse time}, in the LMFP regime ($\kns > 1$) the number of scatterings required to collapse approaches a constant ($\approx200$ as in \cref{eq:tlmfp}), while in the SMFP regime ($\kns \ll 1$) the number of scatterings to collapse grows with the square of the cross section. Note the split marker: two halos at the same Knudsen number have near-identical collapse behaviors.}
    \label{fig:sim_data_const}
\end{figure*}

The number of relaxation timescales required for the constant cross section simulations to reach $t_{\mathrm{col}}$ (as we define it) is shown in \cref{fig:sim_data_const}. For these cases, the collapse time can be well fit by the form:

\begin{equation}
    t_{\mathrm{col}}^{(0)} = (n_{\mathrm{S}} + n_{\mathrm{L}}) \left(1 + A e^{-u^2/(2W^2)}\right) t_{\mathrm{rel}},
    \label{eq:tcolfit}
\end{equation}
where 
\enquote{$(0)$} denotes the velocity-independence of the cross section, $n_{\mathrm{S}} \equiv B\, \kns^{-2}$ is the SMFP number of scatterings to collapse, $n_{\mathrm{L}}$ the LMFP number of scatterings to collapse (a constant $\approx 200$ in \cref{eq:tlmfp}), and $u = \ln(n_{\mathrm{S}}/n_{\mathrm{L}})$. The coefficients $B$, $n_{\mathrm{L}}$, $A$, $W$ are determined by a log-space least-squares fit to the six constant cross section Halo~1 simulations in \cref{tab:halo_parameters}. We find $B= 3.91$, $n_{\mathrm{L}} = 179$, $A=0.917$, and $W=2.23$. The log-normal term in \cref{eq:tcolfit} increases the number of scatterings until collapse in the intermediate MFP regime, which is required to match the simulation data.

However, this model is not accurate for SMFP halos with velocity-dependent cross sections. Recall that the effective cross section (\cref{eq:eff}) requires a well defined $\nu_{\mathrm{eff}}$. In the LMFP regime, the velocity dispersion evolves slowly during the early phases of core formation and collapse, and the thermal conductivity remains near its initial value. In the SMFP regime, this is not the case and the effective cross section ought to properly sample the entire evolutionary history of $\nu_{\mathrm{eff}}$ along the collapse trajectory. For strongly velocity-dependent $\sigma \propto v^{-4}$ cross sections, a factor of few difference in the velocity dispersion is more than an order of magnitude in the cross section. 

We take into account the variation of $\nu$ along the collapse trajectory (i.\,e., the $\rho$-evolution) in the deep SMFP regime as follows. Under the ansatz that each $e$-fold of central density growth takes time $t_\kappa$, we write
\begin{equation}
    \label{eq:ansatz}
    \diff{\ln(\rho_{\mathrm{c}})}{t} = \frac{1}{t_{\kappa}(L_{\mathrm{max}})}, 
\end{equation}
with $L_\kappa$ in \cref{eq:t_kappa} set to the peak luminosity $L_{\mathrm{max}}$. Then, integrating \cref{eq:ansatz} until the collapse criterion $\rho_{\mathrm{c}}/\rho_{\mathrm{c,min}}=25$ is reached, we have 
\begin{equation}
    t_{\mathrm{col}} \approx \int_0^{\ln(25)} t_{\kappa} \dl\ln(\rho_{\mathrm{c}}/\rho_{\mathrm{c, min}}),
\end{equation}
so that in the deep SMFP regime (using \cref{eq:kappa_smfp,eq:t_kappa})
\begin{equation}
    \label{eq:tcolsmfp}
    t_{\mathrm{col}} \approx \frac{1}{G}\int_{0}^{\ln(25)}\Gamma_{\mathrm{eff}}^{(2)}(\nu)\nu \dl\ln(\rho_{\mathrm{c}}/\rho_{\mathrm{c, min}})
\end{equation}
where $\nu$ is evaluated at $L_{\mathrm{max}}(\rho_{\mathrm{c}})$. 

We treat the effect of the velocity dependence of the cross section on the SMFP collapse by introducing the ratio $\Psi$. This is the ratio of \cref{eq:tcolsmfp} for velocity-dependent and velocity-independent cross sections deep in the SMFP regime (at the same Knudsen number),
\begin{equation}
   \Psi \equiv \frac{t_{\mathrm{col}}}{t^{(0)}_{\mathrm{col}}}\Bigg|_{\mathrm{SMFP}}
   \approx\frac{\displaystyle\int{\Gamma_{\mathrm{eff}}^{(2)}\nu} \dl\ln(\rho_{\mathrm{c}}/\rho_{\mathrm{c, min}})}{\displaystyle\frac{44}{45}\tilde{\sigma}_{\mathrm{m}} \int \nu \dl\ln(\rho_{\mathrm{c}}/\rho_{\mathrm{c, min}})}.
    \label{eq:tcol_int}
\end{equation}
$\Gamma_{\mathrm{eff}}^{(2)} = (44/45)\tilde{\sigma}_{\mathrm{m}}$ in the denominator comes from $\Gamma_{p}= \tilde{\sigma}_{\mathrm{m}}$ for $p = \numlist{5; 7; 9}$ in \cref{eq:g2} for velocity-independent cross sections. $\Psi$ can then be used to rescale the SMFP contribution in the calibrated fit in \cref{eq:tcolfit}, as we do below.

We find that the heating trajectory $\nu(\rho_{\mathrm{c}})$ is not very sensitive to the details of the cross section, so the same $\nu(\rho_{\mathrm{c}})$ can be used in both the numerator and the denominator of \cref{eq:tcol_int}. To simplify $\Psi$, we define a weighted average along this common trajectory,
\begin{equation}
    \label{eq:gbar}
    \bar{g} \equiv \frac{\displaystyle \int g \nu \dl\ln(\rho_{\mathrm{c}}/\rho_{\mathrm{c, min}})}{\displaystyle \int \nu \dl\ln(\rho_{\mathrm{c}}/\rho_{\mathrm{c, min}})}.
\end{equation}
With $g=\Gamma_{\mathrm{eff}}^{(2)}$, \cref{eq:tcol_int} becomes
\begin{equation}
    \Psi = \frac{45}{44}\frac{\overline{\Gamma_{\mathrm{eff}}^{(2)}}}{\tilde{\sigma}_{\mathrm{m}}},
\end{equation}
which can be approximated by two-point quadrature. In terms of variables $x = \ln(\nu/\nu_{\mathrm{eff}})$, and $\sigma_x^2 = \overline {(x-\bar{x})^2}$,
\begin{equation}
    \Psi \approx \frac{45}{88} \frac{\left(\Gamma_{\mathrm{eff}}^{(2)}(\hat x + \sigma_x) + \Gamma_{\mathrm{eff}}^{(2)}(\hat x - \sigma_x) \right)}{\tilde{\sigma}_{\mathrm{m}}},
    \label{eq:psiest}
\end{equation}
where we have replaced the exact first moment $\bar{x}$ (i.\,e., \cref{eq:gbar} with $g=x$) with the free (calibrated) parameter~$\hat{x}$.

Then, the collapse time for velocity-dependent cross sections can be written with the SMFP contribution rescaled as $n_{\mathrm{S}} \rightarrow \Psi n_{\mathrm{S}}$,
\begin{equation}
    t_{\mathrm{col}} = (\Psi n_{\mathrm{S}} + n_{\mathrm{L}}) \left(1 + A e^{-(u + \ln(\Psi))^2/(2W^2)}\right) t_{\mathrm{rel}},
    \label{eq:tcolpsi}
\end{equation}
where $n_{\mathrm{S}}$, $n_{\mathrm{L}}$, $A$, $W$, and $u$ are unchanged from the velocity-independent case in \cref{eq:tcolfit}. From the deepest velocity-independent SMFP run (i.\,e., Halo~1, $\tilde{\sigma}_{\mathrm{m}} = \qty{2.5e5}{\cm\squared\per\g}$), we find $\sigma_x = 0.12$, while $\hat x = 0.052$ is calibrated by requiring that \cref{eq:psiest} give $t_{\mathrm{col}}$ exactly for the deepest velocity-dependent SMFP simulation of Halo~1.

\begin{figure*}
    \centering
    \includegraphics[width=\linewidth, height=0.87\textheight, keepaspectratio]{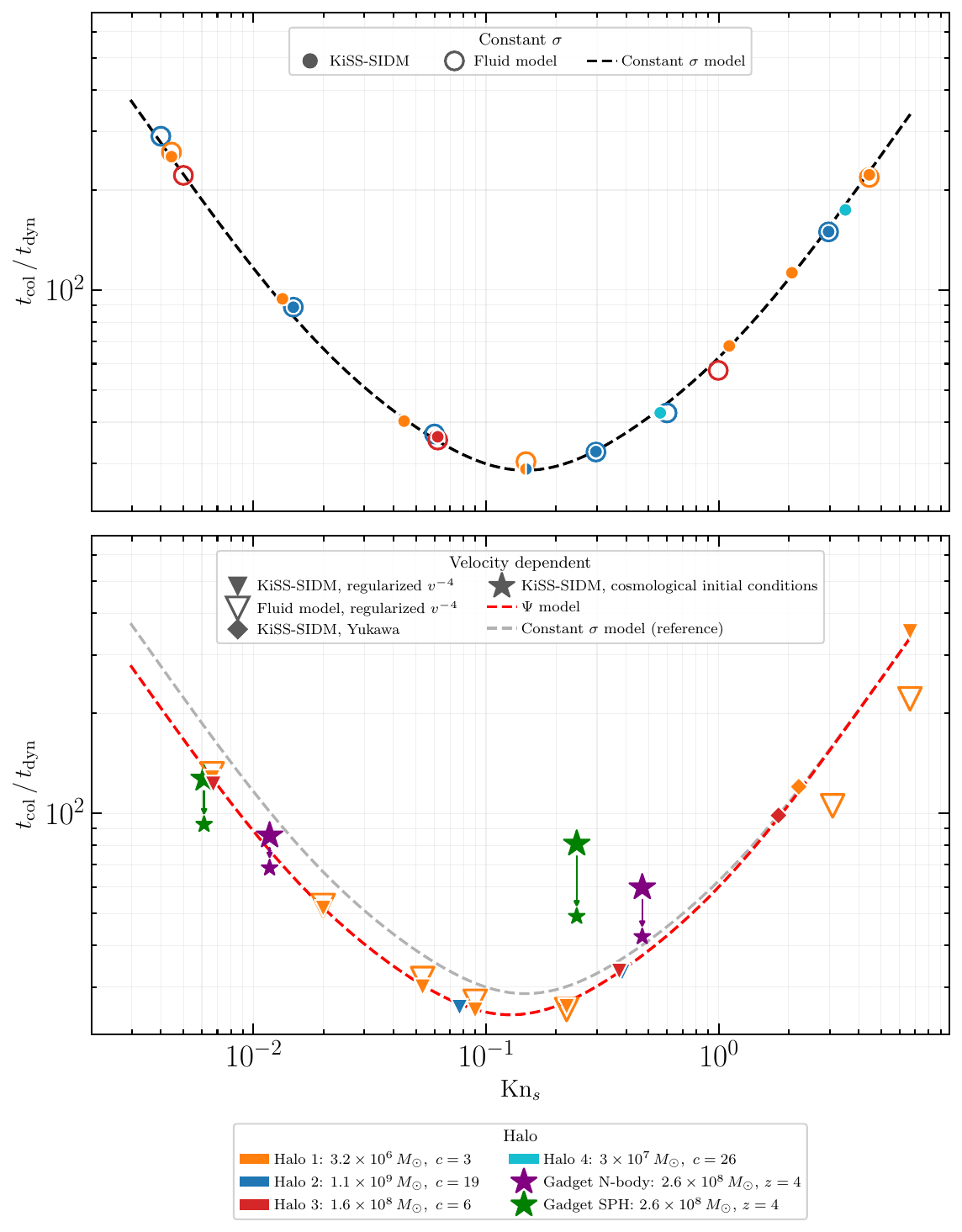}
    \vspace{-11pt}
    \caption{Summary figure of the main results of this work. Shown here is the ratio of the collapse timescale $t_{\mathrm{col}}$ to the dynamical timescale $t_{\mathrm{dyn}} \equiv 1/\sqrt{G\rho_{\mathrm{s}}}$ as a function of the Knudsen number at the scale radius $\kns$ for all simulations in this work: kinetic (\codename) with idealized NFW initial conditions (filled circles, triangles, diamonds), kinetic ({\codename}) with cosmological initial conditions (filled star symbols), and fluid (\textsc{SIDMStars}) with idealized NFW initial conditions (unfilled symbols). All halo properties of these simulations are listed in \crefrange{tab:halo_parameters}{tab:fluid_runs}. The top panel shows the constant cross section simulations, whereas the bottom panel shows the velocity-dependent (regularized $\sigma\propto v^{-4}$ and fully differential Yukawa) ones. Also shown are the fitting formulae (dashed curves) that we identify for each cross section category: \cref{eq:tcolfit} in the top panel and \cref{eq:tcolpsi} in the bottom, with $t_{\mathrm{rel}}\rightarrow (\kns/\sqrt{4\pi}) t_{\mathrm{dyn}}$.}
    \label{fig:tcol_tdyn_all}
\end{figure*}

Expressing the collapse time as a multiple of the relaxation time has the primary advantage that in the LMFP regime, $t_{\mathrm{col}}/t_{\mathrm{rel}} \rightarrow n_{\mathrm{L}}$ becomes constant (as in \cref{eq:tlmfp}; equivalently, $n_{\mathrm{L}} \gg \Psi n_{\mathrm{S}}$ in \cref{eq:tcolpsi}). This advantage is lost in the short  and intermediate MFP regimes when the Knudsen number explicitly enters the collapse time ($\Psi n_{\mathrm{S}} \gtrsim n_{\mathrm{L}}$). Then, it is perhaps clearer to express the collapse time as a multiple of the dynamical timescale via 
\begin{equation}
    \frac{t_{\mathrm{col}}}{t_{\mathrm{dyn}}} = \frac{t_{\mathrm{col}}}{t_{\mathrm{rel}}}\frac{\kns}{\sqrt{4\pi}}, 
\end{equation}
where $t_{\mathrm{dyn}} \equiv 1/\sqrt{G \rho_{\mathrm{s}}} = (\sqrt{4\pi}/\kns)t_{\mathrm{rel}}$. In this form, it is more straightforward to visually compare the collapse timescale for the same halo at different cross sections. The collapse times for all of the {\codename} simulations described so far, as well as the for the cosmological initial conditions and fluid simulations, which will be described in the following sections, are shown in \cref{fig:tcol_tdyn_all}.
The idealized simulations (labeled \enquote{\codename} in the top panel for constant $\sigma$, and \enquote{\codename, regularized $\sigma\propto v^{-4}$} and \enquote{\codename, Yukawa} in the bottom panel) agree to within $\lesssim \qty{10}{\percent}$ of their respective model predictions. 
The dashed curves shown are the aforementioned model predictions: \cref{eq:tcolfit} (black, top panel) and \cref{eq:tcolpsi} (red, bottom panel) with $t_{\mathrm{rel}}\rightarrow (\kns/\sqrt{4\pi}) t_{\mathrm{dyn}}$.

\section{Cosmological Simulations}
\label{sec:comsosim}

The idealized, isolated halos of the preceding section are a gross simplification of realistic cosmological halos. Previous work shows that mergers and accretion can modify the collapse timescale \cite{Elbert_2015, Silverman:2026jvu}. Of greater concern for the present work, the NFW profile arises from hierarchical structure formation only in collisionless dark matter, and so our assumed NFW initial conditions may not be appropriate for collisional dark matter. In the LMFP and intermediate-MFP regimes, the NFW initial conditions adopted in \cref{sec:kin_sim} may be reasonable. However, in the SMFP regime, a typical particle will have experienced many scatterings during the hierarchical assembly of present-day dwarf galaxy-mass halos. Further, accretion is likely to differ qualitatively for strongly collisional dark matter; in this regime, one might expect infalling material to shock at the halo boundary, with some associated entropy production. As an initial exploration of the extent to which these effects may modify the core collapse time, and to motivate further work, we run cosmological simulations in the strongly collisional and collisionless limits. From these we extract \enquote{less unrealistic} halos (as compared with the isolated NFW assumption of the preceding section), which we evolve with {\codename}.

\subsection{Simulation Setup}

Using the code \textsc{Gadget-4} \cite{Springel_2021}, we run matched $N$-body (collisionless DM) and smoothed particle hydrodynamics (SPH; strongly collisional fluid limit) simulations  from the same initial particle configuration, so that differences between the resulting halos reflect only the collisional physics. The SPH run approximates the strongly collisional (${\mathrm{Kn}}\to0$) limit of SIDM, in which the dark matter behaves as an ideal fluid that can shock but does not conduct heat. Specifically, both $N$-body and SPH simulations use the same $256^3$ N-GenIC \cite{ngenic_2015} realization of a \qty{0.35}{\per\hHubble\Mpc} comoving periodic box, started at $z=159$ and evolved to $z=4$. The cosmology is Planck~2018 \cite{Planck_2018}. The SPH run has $64$ neighbors, and an artificial viscosity parameter of unity. The gravitational softening length is $\epsilon = 50.7$ comoving \unit{pc}, i.\,e.\ $10.2$ physical \unit{pc} at $z = 4$.

From these simulations, we extract a sphere of radius $3.8\,r_{200\mathrm{c}} \approx 15 \text{ physical \unit{\kpc}}$ around a halo at redshift $z=4$. The cutout contains \num{1.77e6} (\num{1.76e6}) particles for the $N$-body (SPH) run, of which \qty{45}{\percent} lie inside $r_{200\mathrm{c}}$ for both halos. This halo has a mass at $r_{200\mathrm{c}}$ of \qty{2.63e8}{\Msun}
(\qty{2.61e8}{\Msun} for the SPH run), with
$r_{200\mathrm{c}} = 3.96 \text{ physical \unit{\kpc}}$ and $v_{\mathrm{max}} = 18.7$ ($18.5$) \unit{\km\per\s}.

\begin{figure}
    \centering
    \includegraphics[width=\linewidth]{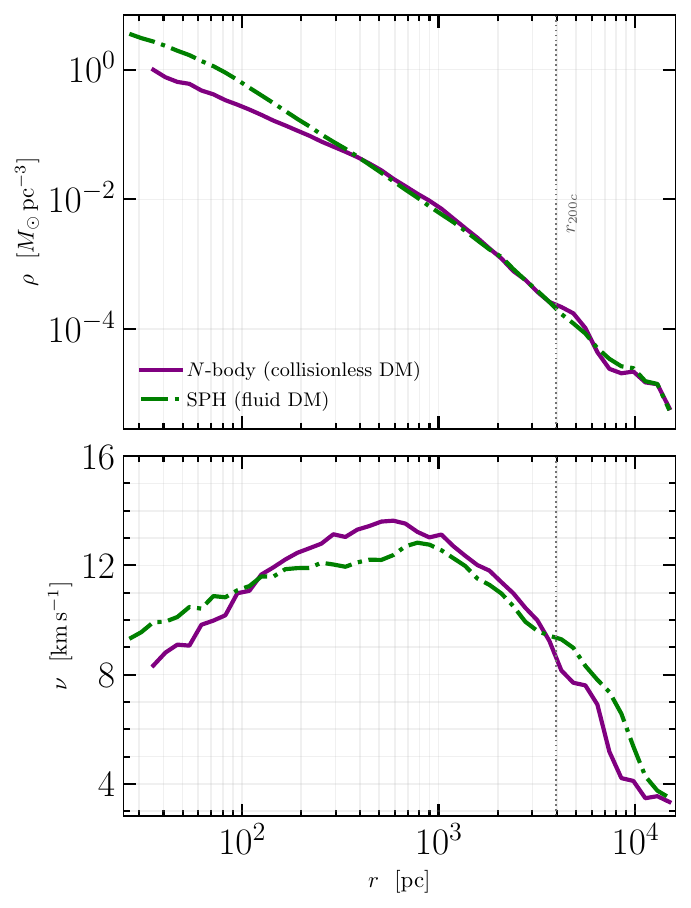}
    \caption{Spherically averaged density (top) and one-dimen\-sional velocity dispersion
    (bottom) of the $z=4$ halo cutouts, for the $N$-body (solid purple) and SPH (dash-dotted green) simulations. The SPH dispersion includes the thermal support, calculated from the internal energy of the particles. The vertical dotted line marks $r_{200\mathrm{c}} \approx \qty{4000}{\pc}$. The two halos agree reasonably well between $\qty{300}{\pc} \lesssim r \lesssim \qty{4000}{\pc}$; inside $r \approx \qty{300}{\pc}$, the SPH halo is denser and steeper.}
    \label{fig:cosmo_profs}
\end{figure}

We spherically average these cutout regions, which we use to initialize the kinetic {\codename} collapse simulations at two cross sections, in the transition regime and the deep SMFP regime.
The spherically-averaged density and velocity dispersion profiles of the cutouts at $z=4$ are shown in \cref{fig:cosmo_profs}. Although the profiles agree reasonably well at larger radii, the central region is denser in the SPH runs.
The properties and parameters of these simulations are recorded in \cref{tab:gadget_parameters}. At the end of the cosmological simulation ($z=4$), the halos are still accreting rapidly. This accretion is preserved in the kinetic simulations, but is symmetrized.

Comparing these simulations with the model prediction (\cref{eq:tcolpsi}) requires calculating a Knudsen number and velocity scale. The former is supplied by fitting an NFW profile to the density profile out to  $0.5 r_{200\mathrm{c}}$. Meanwhile, $v_{\mathrm{max}}$ (which sets $\tilde{\sigma}_{\mathrm{m}}$ through \cref{eq:eff}) is determined directly from the rotation curve.

\subsection{The Collapse Time with Cosmological Initial Conditions}

\begin{figure}
    \centering
    \includegraphics[width=\linewidth]{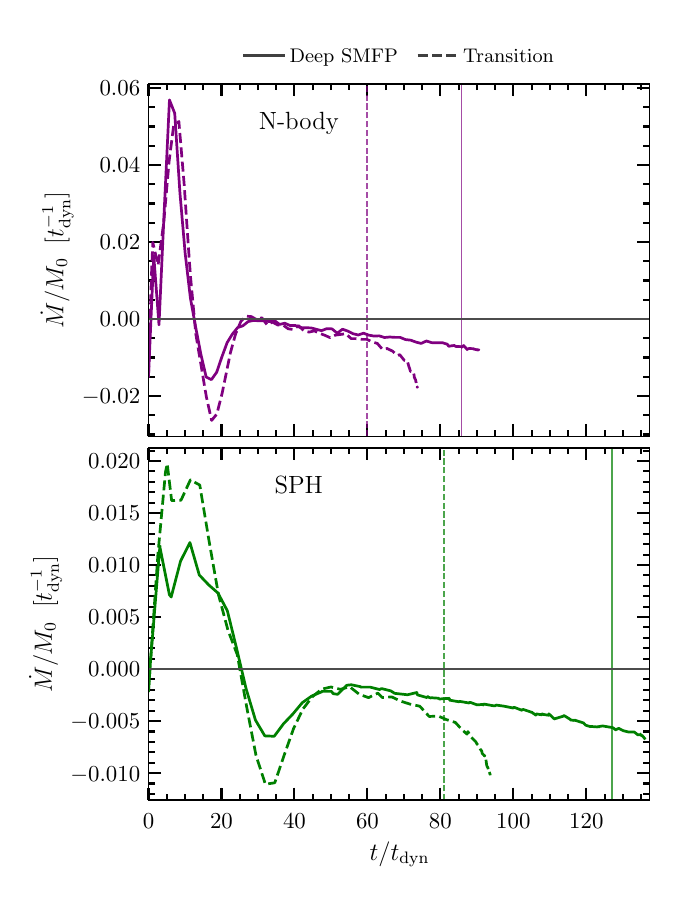}
    \caption{Infall rate at $r_{200\mathrm{c}}$ for the kinetic simulations initialized from $N$-body (top, purple) and SPH (bottom, green), in the deep SMFP ($\tilde{\sigma}_{\mathrm{m}} \approx \qty{1.7e4}{\cm\squared\per\g}$, $\kns\lesssim0.01$, solid curves) and transition ($\tilde{\sigma}_{\mathrm{m}} \approx \qty{4e2}{\cm\squared\per\g}$, $0.3 < \kns < 0.5$, dashed curves) regimes. An initial bulk infall ($\dot{M}/M_{200\mathrm{c}} > 0$) is followed by a weaker splashback (a dip down to $\dot{M}/M_{200\mathrm{c}} \approx -0.01$), before settling into a bulk outflow consistent with an evaporative wind. The collapse times are indicated by vertical lines (corresponding to collapse times marked by the larger star symbols in \cref{fig:tcol_tdyn_all}).}
    \label{fig:mdot}
\end{figure}

We find that the infalling material delays the collapse (increases $t_{\mathrm{col}}$) compared to the idealized runs of \cref{sec:kin_sim}.  
The collapse times are shown in \cref{fig:tcol_tdyn_all} (star symbols).
The accretion rate at $r_{200\mathrm{c}}$ in the {\codename} simulations is shown in \cref{fig:mdot}, in which we see that an initial bulk infall is followed by a weaker splashback, before settling into a bulk outflow consistent with an evaporative wind. 

As an attempt to build a model that better predicts collapse times for cosmological halos, we control for accretion by subtracting from $t_{\mathrm{col}}$ the time elapsed before the peak of the \enquote{splashback}. The splashback time $t_{\mathrm{sb}}$ is taken to be where $\dot{M}/M_0$ in \cref{fig:mdot} dips to a local minimum $\approx -0.01\,t_{\mathrm{dyn}}^{-1}$, before rising again.

This correction $t_{\mathrm{col}}\to t_{\mathrm{col}}-t_{\mathrm{sb}}$ is indicated by the downward arrows and smaller star symbols in \cref{fig:tcol_tdyn_all}.

After this (ad hoc) correction, the $N$-body halos fall within $\approx \qty{15}{\percent}$ of the model-predicted collapse time. However, for the SPH case, the model is accurate only to a factor of two to three. Of course, neither the $N$-body nor the SPH simulations are correct treatments of the SIDM dynamics. Further work is needed quantify the role of cosmological evolution on the collapse time, especially in the short and intermediate MFP regimes.

\begin{table*}
    \centering
    \caption{Summary of halo properties, scattering cross sections, and particle numbers of the kinetic (\codename) simulations started from cosmological initial conditions (\cref{sec:comsosim}).
    Both halos are the same $z=4$ object taken from matched Gadget $N$-body (collisionless) and SPH simulations, extracted out to $3.8\,r_{200\mathrm{c}}$.
    The mass is measured at $z = 4$ and $c = r_{200\mathrm{c}}/r_{\mathrm{s}}$. $\rho_{\mathrm{s}}$ and $r_{\mathrm{s}}$ come from NFW fits out to $0.5\,r_{200\mathrm{c}}$, while
    $\tilde{\sigma}_{m}$ is the effective cross section (\cref{eq:eff}) evaluated at $\nu_{\mathrm{eff}} = 0.71\,v_{\max}$.}
    \smallskip
    \setlength{\tabcolsep}{5pt}
    \begin{tabular}{l c *{2}{S[table-format=1.2e1]} c S[table-format=1.2e+1] c S[table-format=1.2e1] S[table-format=1.2e+1] l S[table-format=1.2e1]}
    \toprule
    Halo & {$z$} & {$M_{200\mathrm{c}}$} & {$M_{\mathrm{tot}}$} & {$c$} & {$\rho_{\mathrm{s}}$} & {$r_{\mathrm{s}}$} & {$\tilde{\sigma}_{\mathrm{m}}$} & {$\mathrm{Kn}_s$} & Type & {$N$} \\
    & & {(\unit{\Msun})} & {(\unit{\Msun})} & & {(\unit{\Msun\per\pc\cubed})} & {(\unit{\pc})} & {(\unit{\cm\squared\per\g})} & & & \\
    \midrule
    \textsc{Gadget} & 4 & 2.63e8 & 5.87e8 & 5.05 & 4.80e-2 & 784
        & 2.71e2 & 4.69e-1 & regularized $v^{-4}$  & 1.77e6 \\
    ($N$-body) & & & & & & & 1.08e4 & 1.17e-2 & regularized $v^{-4}$  & 1.77e6 \\

    \addlinespace

    \textsc{Gadget} & 4 & 2.61e8 & 5.85e8 & 8.74 & 1.53e-1 & 452
        & 2.81e2 & 2.45e-1 & regularized $v^{-4}$  & 1.76e6 \\
    (SPH) & & & & & & & 1.13e4 & 6.13e-3 & regularized $v^{-4}$  & 1.76e6 \\

    \addlinespace

    \bottomrule
    \end{tabular}
    \label{tab:gadget_parameters}
\end{table*}

\section{Fluid Model: \textnormal{\textsc{SIDMStars}}}
\label{sec:fluid}

Although the kinetic simulations of the previous section are not enormously expensive (hours to days on consumer hardware depending on the particular parameters), the cost is prohibitive for exploring very large parameter spaces. Further, time step and resolution requirements limit the {\codename} simulations to following several decades in central density evolution. At the conclusion of these runs, it is not necessarily clear whether the evolution has entered a self-similar phase that could be extrapolated to still higher central densities. These considerations have motivated the development of an alternative computational framework, the so-called \enquote{gravothermal fluid model} \cite{Lynden-Bell80, Balberg_2002, Nishikawa_2020, Outmezguine:2022bhq, Gad-Nasr:2023gvf}. This approach involves solving the four coupled differential equations describing the structure of a spherically symmetric, hydrostatic halo with an ideal gas equation of state. Namely,
\begin{align}
    &\diffp{r}{m} = \frac{1}{4\pi r^2 \rho} && \text{(mass continuity)} \\
    \displaybreak[0]
    &\diffp{P}{m} = -\frac{Gm}{4\pi r^4} && \text{(hydrostatic equilibrium)}
    \label{eq:hse} \\
    \displaybreak[0]
    &L_\kappa = -24\pi^2r^4\rho \kappa \diffp{\nu^2}{m} && \text{(Fourier's law for conduction)}
    \label{eq:conduction} \\
    \displaybreak[0]
    &\diffp{L_\kappa}{m} = -\nu^2 \diffp{s}{t} && \text{(energy conservation)}
\end{align}
where $m$ is the mass enclosed within radius $r$, $P=\rho\nu^2$ is the pressure at mass coordinate $m$, $L_\kappa$ is the conductive luminosity through the sphere enclosing mass $m$ (first introduced in \cref{eq:t_kappa}), $\kappa$ is the conductivity defined in \cref{eq:q}, and $s = \ln(\nu^3/\rho)$ is the specific entropy (up to an additive constant). The prefactor $24$ in the luminosity equation arises from combining the geometric $16 \pi ^2$ with Fourier's law and $u = (3/2) \nu^2$.

In the SIDM literature, these equations are usually solved via alternately evolving the thermal structure (through finite-difference updates of the internal energy $u=(3/2)\nu^2$ in each cell) and the mechanical structure (by iteratively solving a linearized form of the hydrostatic equilibrium condition, \cref{eq:hse}), e.\,g.\ in \citet{Balberg_2002}. This method is eventually limited by the slow convergence of the hydrostatic adjustment in the deeply collapsed regime. Formally, the fluid model is identical to the equations of stellar structure, except that the luminosity is conductive. In that context, the modern approach is a fully implicit Henyey method, which simultaneously evolves all four discretized structure equations \cite{Henyey_1964, Kippenhahn2012}.

\subsection{\textnormal{\textsc{SIDMStars}}: Numerical Scheme}

An alternative to the Henyey method, which also enjoys a long history in the stellar structure literature and has the advantage of numerical simplicity, is a shooting approach. We implement a fluid code using this technique, which we call.\textsc{SIDMStars}%
\footnote{Publicly available at:\\\url{https://github.com/jamesgurian/SIDMStars}.}

Specifically, the approach we adopt is the \enquote{explicit} method of \cite{Stahler1980b, Hosokawa2009}. As in the \citet{Balberg_2002} method, the hydrostatic structure and the energy transport are solved separately. The hydrostatic structure (at fixed $s(m)$) is solved as a boundary value problem with one inner and one outer boundary condition. For the outer boundary condition we assume fixed surface pressure $P_{\mathrm{surf}}$, while for the inner boundary condition we assume homologous evolution of the innermost region:
\begin{align}
    &P(M_{\mathrm{max}}) = P_{\mathrm{surf}} && \text{ (outer boundary)}
    \label{eq:P_outer_bc}
    \\
    &\ln\mleft(\frac{r_{\mathrm{in}}^{i+1}}{r_{\mathrm{in}}^i}\mright) = - \frac{1}{4} \ln\mleft(\frac{P_{\mathrm{in}}^{i+1}}{P_{\mathrm{in}}^i}\mright),~ && \text{(inner boundary)}
\end{align}
where the subscript \enquote{$\mathrm{in}$} indicates the innermost unresolved central sphere of mass $m_{\mathrm{in}}$ (we do not integrate from $m=0$ due to the singularity in the continuity equation in these variables) and the superscripts indicate the time step. Then, using guessed values of $P_{\mathrm{in}}^{i+1}$ and $r_{\mathrm{max}}$ the structure equations are integrated outwards from the center and inwards from the edge using \texttt{DifferentialEquations.jl} \cite{rackauckas2017differentialequations}. At a fitting mass (which is adaptively adjusted to ensure that both integrations reach the fitting point), the values of $P$ and $r$ generated by the two integrations are compared and used to produce improved guesses. That is, the boundary value problem becomes a nonlinear solve, which we treat via the Julia package \texttt{NonlinearSolve.jl} \cite{pal2024nonlinearsolve}. We adopt this approach because the boundary value problem interface provided with \texttt{DifferentialEquations.jl} struggles with the sensitivity to the initial guesses in this problem, which tends to cause trial solutions to diverge. 

Then, at fixed hydrostatic (mechanical) structure we explicitly update the entropy by finite differencing $\nu^2$ and $L_\kappa$ to determine $\difsp{s}{t}$. Again, there are two boundary conditions. For the central luminosity boundary condition we take:
\begin{equation}
    L_{\mathrm{in}} = -\frac{E_{\mathrm{in}}^{i+1}-E_{\mathrm{in}}^i}{\Delta t},~~~~~~~ \text{ (inner boundary)}
\end{equation}
where the energy $E_{\mathrm{in}}$ is the total (thermal plus gravitational) energy of the central sphere of mass $m_{\mathrm{in}}$ at successive mechanical structures (time steps $i$ and $i+1$). $E_{\mathrm{in}}$ is calculated from the central density, radius, and pressure by assuming a uniform-density spherical center interior to $m_{\mathrm{in}}$. For the outer boundary condition, meanwhile, we simply take 
\begin{equation}
    L_\kappa(M_{\mathrm{max}}) = 0 ~~~~~~~~~~~ \text{ (outer boundary)}
\end{equation}
i.\,e., no heat flows out of the halo. We have verified that instead fixing the entropy at the outer edge, $\diffp{s}{t} \big|_{M_{\mathrm{max}}} \!\! = 0$, produces similar results.

The entropy update has the structure of a diffusion problem, and the time step is thus restricted by a Courant-Friedrichs-Lewy (CFL) condition:
\begin{equation}
    \label{eq:dt_cfl}
    \Delta t_{\mathrm{CFL}} \lesssim \frac{\Delta m^2}{16 \pi^2 r^4 \rho \kappa}, 
\end{equation}
where $\Delta m$ is the mass of a shell, and the condition is evaluated in every shell, with the minimum setting the time step.
This is generally much more restrictive than the thermal (Kelvin-Helmholtz) timescale on which the hydrostatic structure evolves. Thus, we sub-cycle the entropy update on a user-configurable fraction of the most restrictive CFL limit. The hydrostatic structure is updated on a user-configurable but small fraction of the smallest thermal timescale, $t_{\mathrm{therm}} = 1/|\diffp{s}{t}|$.


\subsection{Conductivity Interpolation}
Commonly in the fluid treatment of SIDM in the literature, the conductivity is given as 
\begin{equation}
    \label{eq:kform_std}
    \kappa_{\mathrm{fluid}} \equiv \frac{1}{\kappa_{\mathrm{SMFP}}^{-1}+ \kappa_{\mathrm{LMFP}}^{-1}},
\end{equation}
a harmonic interpolation between the two MFP regimes. Recall that $\kappa_{\mathrm{SMFP}}$ and $\kappa_{\mathrm{LMFP}}$ were defined in \cref{eq:kappa_smfp,eq:kappa_lmfp}. However, we find that, in the intermediate MFP regime, this choice of model for the conductivity produces shorter fluid collapse times than the kinetic simulations of the previous sections. This motivates a form for the conductivity that is additionally suppressed relative to \cref{eq:kform_std} and slows the collapse (increases $t_{\mathrm{col}}$) in the intermediate MFP. Inspired by the model fit to the collapse times (\cref{eq:tcolfit}), we adopt the form 
\begin{equation}
    \tilde\kappa_{\mathrm{fluid}} \equiv \frac{\kappa_{\mathrm{fluid}}}{1+ \tilde{A} \exp\mleft(-y^2/(2\tilde{W}^2)\mright)},
    \label{eq:kform}
\end{equation}
with $y \equiv \ln\mleft(\left.\kappa_{\mathrm{LMFP}}\right/\kappa_{\mathrm{SMFP}}\mright)$.
In total, we have three free parameters in \cref{eq:kform}: $\tilde{A}$, $\tilde{W}$, and the parameter $C$ in $\kappa_{\mathrm{LMFP}}$. We first fix $C$ by iterating on fluid simulations matched to the deepest LMFP kinetic run of Halo~1. In the deep LMFP regime, the collapse time is nearly linear in $1/C$, and we converge to $C=0.85$. For $\tilde A$ and $\tilde W$, we are able to measure (in aggregate) the heat flux directly from the kinetic simulations in bin $j$:
\begin{equation}
    \label{eq:q_recon}
    q_j = \frac{1}{V_j}\sum_{n \in j}\frac{1}{2}\, m_n\, |\vec{v}_n|^2\, v_{r,n},
\end{equation}
where the sum is over the particles $n$ in bin $j$, $V_j=4\pi(r_j^3-r_{j-1}^3)/3$ is the volume of the bin, $m_n$ is each particle's mass (the same for all $n$), $\vec{v}_n$ is the peculiar (bin-mean-subtracted) velocity, and $v_{r,n}$ is its radial component.

The data are smoothed temporally by fitting a quadratic to $\ln(r)$, $\ln(\nu^2)$, and $\ln(\rho)$ over the preceding and following snapshot and spatially by fitting a quadratic in the same quantities to the two neighboring spatial bins. Then, we can construct the conductivity in each bin,
\begin{equation}
    \label{eq:kappa_from_q_recon}
    \kappa_j = - \frac{q_j}{\left.\difs{u}{r}\right|_j}. 
\end{equation}
This estimate may be biased if the variations are not well-described by quadratics. Nevertheless, we perform a least-squares fit of \cref{eq:kform} to three intermediate constant cross section kinetic simulations of Halo~1 collectively ($\tilde{\sigma}_{\mathrm{m}} \in \{\numlist[list-final-separator=\text{,}\ ]{7.48e3; 2.50e4; 8.33e4}\} \, \unit{\cm\squared\per\g}$; \cref{tab:halo_parameters}). Each bin from each snapshot contributes an inverse-variance weighted $\kappa_j$. For each snapshot, bins are constructed targeting $\Delta y \approx 0.4$, each containing at least \num{8000} particles. The first bin is omitted from the fit (although this has negligible effect on the best fit parameters). Bins with $|y|>6$ are cut because they are deep in the LMFP or SMFP regimes where $\tilde\kappa_{\mathrm{fluid}}\to\kappa_{\mathrm{fluid}}$, and contain negligible information about $\tilde{A}$ or $\tilde{W}$. We also exclude bins with Mach number $\mathcal{M} > 0.1$, because in these areas the distribution function is not locally near-Maxwellian \cite{Gurian_2025} and including them would bias the fit of \cref{eq:kform}.

\begin{figure}
    \centering
    \centerline{\includegraphics[width=1.05\linewidth]{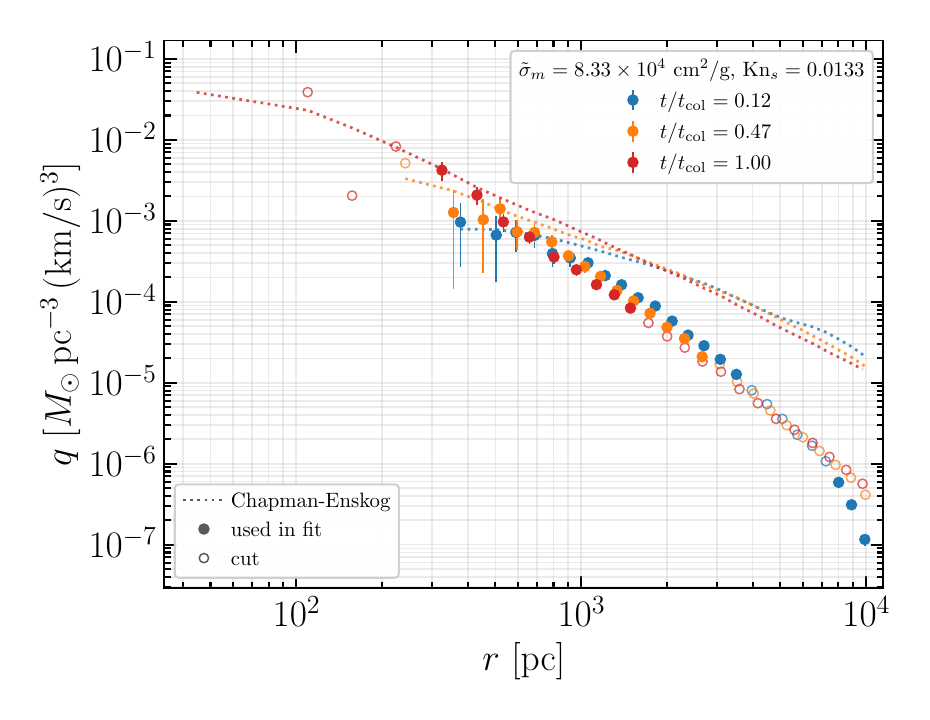}}
    \caption{The reconstructed heat flux $q$ (\cref{eq:q_recon}) at three stages of the collapse in the Halo~1 run with $\tilde{\sigma}_{\mathrm{m}} = \qty{8.33e4}{\cm\squared\per\g}$. The filled-in points are used to fit for parameters $\tilde{A}$ and $\tilde{W}$ in \cref{eq:kform}, while the hollow points are excluded from the fit due to the $|y|$ or the Mach number criteria mentioned in the text below \cref{eq:kappa_from_q_recon}. The dotted curves show the Chapman–Enskog (i.\,e., SMFP regime) prediction for each snapshot. The fact that the central regions (with low $\mathrm{Kn}$) approach this prediction lends credibility to the reconstruction at larger radii (with higher $\mathrm{Kn}$) where no first-principles prediction exists.}
    \label{fig:q_recon}
\end{figure}

As an example, the reconstructed heat fluxes with their Poisson errors of three snapshots from the kinetic Halo~1 run ($\tilde{\sigma}_{\mathrm{m}} = \qty{8.33e4}{\cm\squared\per\g}$) are shown in \cref{fig:q_recon}. This run is the most collisional of the ones used in the calibration of $\tilde A,\ \tilde W$, and especially develops a bulk evaporative wind which eliminates the outer region of the halo from the fit via the cut on $\mathcal{M}$.

This procedure for obtaining $\tilde{A}$ and $\tilde{W}$ is not necessarily statistically or physically optimal: the simulations have different numbers of snapshots and reach different collapse epochs before termination. Nevertheless, it succeeds in calibrating the fluid model to within $\approx \qty{10}{\percent}$ in the velocity-independent case. The best-fit parameters are $\tilde{A} = 1.31$, $\tilde{W} = 1.68$.

For the regularized $v^{-4}$ cross section, the direct application of \cref{eq:kform} with these parameters leads to a $\gtrsim50\%$ underestimation of the collapse time in the LMFP. The choice of the $\Gamma_5$ weighting in \cref{eq:kappa_lmfp} (with $p=5$ in \cref{eq:Gammap}) is not justified by any rigorous derivation, but has been found to work reasonably well in previous work \cite{Outmezguine:2022bhq, Yang_2023}. Here, we find that at the cost of an additional free parameter $f$, the LMFP collapse time is better reproduced by the choice 
\begin{equation}
\label{eq:modif_kappa_lmfp}
   \tilde{\kappa}_{\mathrm{LMFP}} = \frac{aC}{4\pi G}\big[f\,\Gamma_5(\nu) + (1-f)\,\Gamma_7(\nu)\big]\rho\,\nu^3,
\end{equation}
with $f=0.73$---essentially replacing $\Gamma_5$ in \cref{eq:kappa_lmfp} with the combination above. However, this improved fit in the LMFP comes at the cost of overestimating the collapse time in the intermediate MFP by $\approx \qty{30}{\percent}$. 
In principle, the transition weight $f$ could itself be a function of Kn to preserve accuracy in both regimes, but we do not pursue this here. Since the improvement is a trade-off between LMFP and intermediate MFP, we adopt the simpler fiducial choice of $f=1$ in all our plots.

\begin{table}
    \caption{Fluid simulations.
    All runs use $C = \num{0.85}$, $\tilde{A} = \num{1.31}$, $\tilde{W} = \num{1.68}$, and \num{200} grid points.
    $\kns = (r_{\mathrm{s}}\rho_{\mathrm{s}}\tilde{\sigma}_{\mathrm{m}})^{-1}$, $\tilde{\sigma}_{\mathrm{m}}$ is evaluated at $\nu_{\mathrm{eff}} = 0.71\,v_{\max}$ for the velocity-dependent runs.}
    \centering
    \smallskip
    \setlength{\tabcolsep}{5pt}
    \begin{tabular}{l l *{2}{S[table-format=1.2e+1]}}
    \toprule
    Halo & Type & {$\tilde{\sigma}_{\mathrm{m}}$} & {$\kns$} \\
     & & {(\unit{\cm\squared\per\g})} & \\
    \midrule
    \textbf{Halo 1} & regularized $v^{-4}$ & 1.67e2 & 6.65e0 \\
     & constant & 2.50e2 & 4.44e0 \\
     & regularized $v^{-4}$ & 3.59e2 & 3.09e0 \\
     & regularized $v^{-4}$ & 5.00e3 & 2.22e-1 \\
     & constant & 7.48e3 & 1.48e-1 \\
     & regularized $v^{-4}$ & 1.24e4 & 8.98e-2 \\
     & regularized $v^{-4}$ & 2.08e4 & 5.34e-2 \\
     & regularized $v^{-4}$ & 5.57e4 & 1.99e-2 \\
     & regularized $v^{-4}$ & 1.67e5 & 6.65e-3 \\
     & constant & 2.50e5 & 4.44e-3 \\
    \addlinespace
    \textbf{Halo 2} & constant & 5.00e1 & 2.97e0 \\
     & constant & 2.47e2 & 6.00e-1 \\
     & constant & 5.00e2 & 2.97e-1 \\
     & constant & 2.47e3 & 6.00e-2 \\
     & constant & 1.00e4 & 1.48e-2 \\
     & constant & 3.71e4 & 4.00e-3 \\
    \addlinespace
    \textbf{Halo 3} & constant & 1.33e2 & 9.96e-1 \\
     & constant & 2.14e3 & 6.20e-2 \\
     & constant & 2.65e4 & 5.00e-3 \\
    \bottomrule
    \end{tabular}
    \label{tab:fluid_runs}
\end{table}

The collapse times for all fluid simulations (collectively summarized in \cref{tab:fluid_runs}) are shown in \cref{fig:tcol_tdyn_all}, marked by unfilled symbols.
For different concentrations and velocity dependencies, the fluid collapse times agree reasonably well with the kinetic simulations run with idealized NFW initial conditions (\cref{sec:kin_sim}), with the worst agreement for the velocity-dependent deep LMFP cases, as discussed above (see text around \cref{eq:modif_kappa_lmfp}).

\subsection{Convection}
\label{sec:convec}

Late in the collapse, the procedure described above can sometimes produce an entropy inversion, 
\begin{equation}
    \diff{s}{r} < 0.
    \label{eq:schwarz}
\end{equation}
This is the Schwarzschild condition for convective instability. Thus, following the \enquote{equal area} prescription of \cite{Stahler1980b} we have implemented adiabatic convection in our code. That is, before updating the mechanical structure we set the entropy gradient to zero in regions where it has drifted negative, and require that across the convective zone bounded by $m_1,\ m_2$
\begin{equation}
    \int_{m_1}^{m_2} \left[s(M,t)-s_0(t)\right] \nu^2(M, t) \dl M = 0,
\end{equation}
where $s(M,t)$ is the unstable, inverted entropy profile produced by the explicit update, and $s_0(t)$ is the uniform entropy that replaces it.
Since $\nu^2\Delta s$ is heat per unit mass, this condition imposes that the net heat exchange is zero within the convective zone while the entropy gradient is flattened.

\begin{figure*}
    \vspace*{-6pt}

    \centering
    \includegraphics[width=\linewidth, height=0.42\textheight, keepaspectratio]{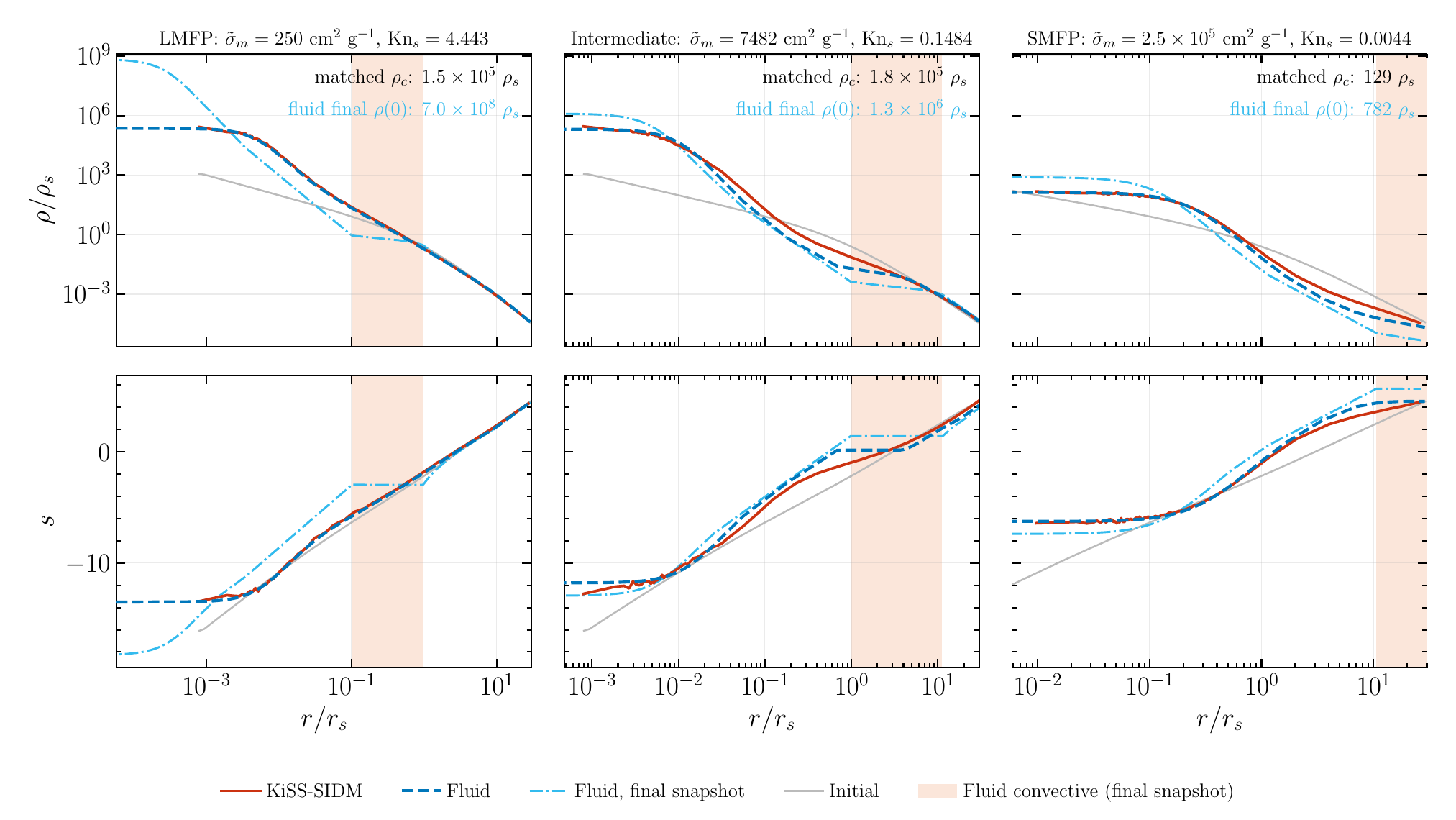}
    \vspace{-15pt}
    \caption{Comparison of fluid (\textsc{SIDMStars}) and kinetic ({\codename}) simulations of Halo~1 (defined in \cref{tab:halo_parameters}) for constant (velocity-independent and isotropic) cross sections. The fluid run is shown in (\textit{i}) dashed dark blue at a snapshot where its central density matches a late-time snapshot of the corresponding {\codename} run in solid red, and (\textit{ii}) dash-dot cyan at its own final snapshot. The convective zone ($\difs{s}{r} < 0$) in the final fluid snapshots is indicated by the shaded vertical band, which develops in the transition region where $\mathrm{Kn}(r) \approx 1$. In the kinetic simulations, this plateau is smoothed out. The convective instability does not seem to form in the velocity-dependent cross section runs, shown in \cref{fig:struct_v4} below. See \cref{sec:convec} for more discussion.}
    \label{fig:conv_struct}

    \vspace{0.2ex}
    \includegraphics[width=\linewidth, height=0.42\textheight, keepaspectratio]{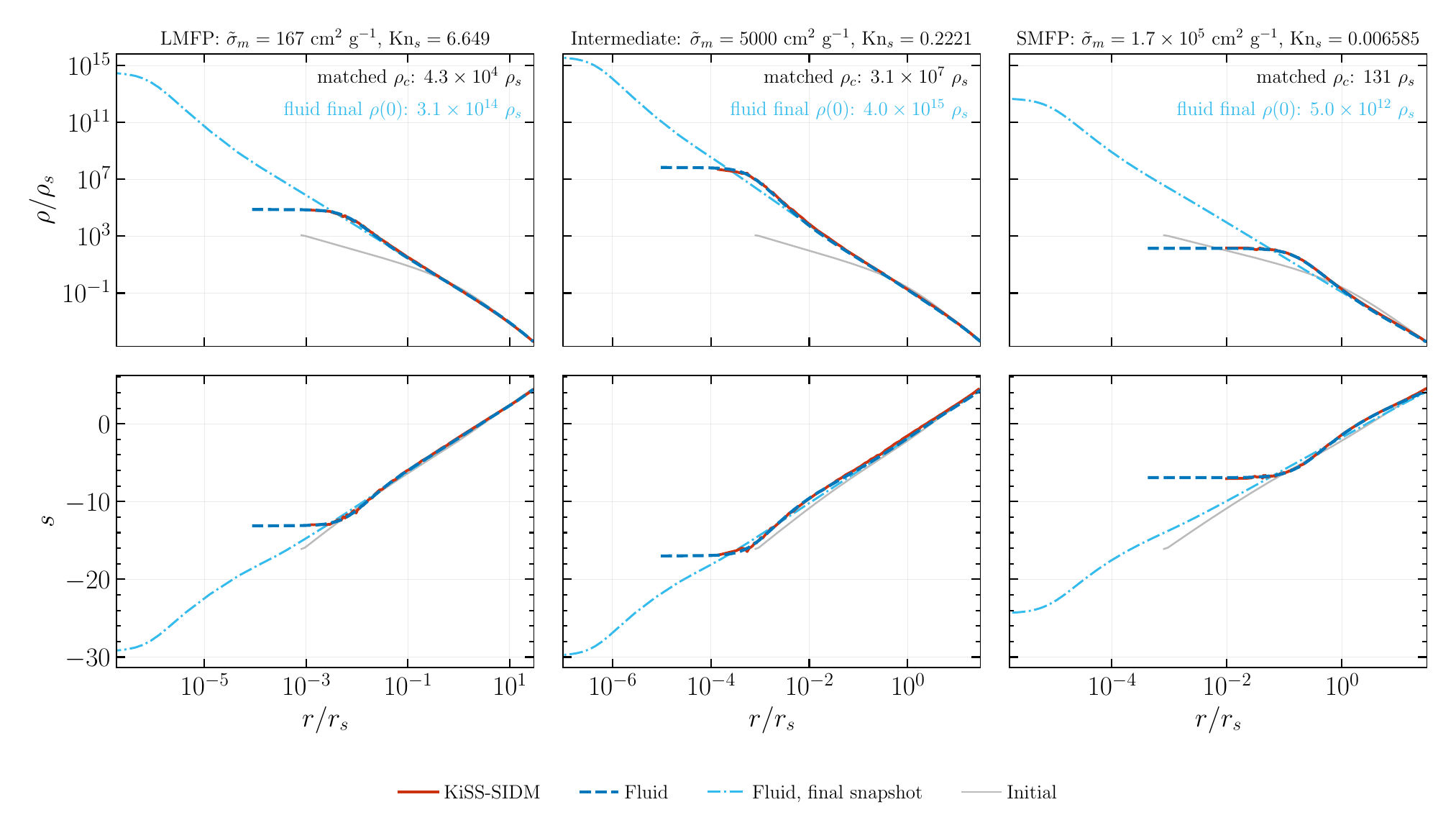}
    \vspace{-15pt}
    \caption{Same as \cref{fig:conv_struct}, but for regularized $v^{-4}$ cross sections. Here, the fluid simulations are able to proceed much deeper into the collapse before the boundary value problem becomes poorly conditioned. At the final {\codename} snapshot, the fluid model agrees well with the kinetic simulations, and does not show any convective zones ($\difs{s}{r} < 0$) like in the constant cross section case (\cref{fig:conv_struct} above).}
    \label{fig:struct_v4}
\end{figure*}

We have found that convection eventually develops in all the halos we simulated with constant (velocity-independent and isotropic) cross sections, around the transition zone (i.\,e., $\mathrm{Kn}(r) \approx 1$) within each halo. The density and specific entropy profiles for three constant cross section {\codename} and fluid simulations are shown in \cref{fig:conv_struct}. In the LMFP (leftmost panels) the results agree well at the last {\codename} snapshot. In the other two cases (middle and rightmost panels), the profiles disagree outside the core. 

Only the intermediate ($\tilde{\sigma}_{\mathrm{m}} = \qty{7.48e3}{\cm\squared\per\g}$; middle panels in \cref{fig:conv_struct}) kinetic simulation was run to the point where convection is established in the fluid runs. 
At that point in the evolution, the kinetic specific entropy profile (red curve) is not flat or decreasing outward within the convective zone that has developed in the corresponding fluid run (dashed and dash-dotted blue curves).
In the transition regime ($\mathrm{Kn}(r) \approx 1$) within the halo, heat is carried by ballistic motions over the temperature gradient scale, which cannot be captured by the hydrostatic fluid treatment. Yet, with only one spatial dimension {\codename} could not properly represent convective overturn.

In any case, this phenomenology is of limited immediate interest, because for velocity-dependent cross sections, we do not find convective zones for any of the parameters studied here. In fact, for the regularized $v^{-4}$ cross sections the kinetic and fluid results agree quite closely, as shown in \cref{fig:struct_v4}. Further, for these velocity-dependent cross sections the fluid simulations are able to advance much deeper into the collapse before the error tolerance on the central pressure becomes smaller than 64-bit machine precision.

\section{Conclusions and Discussion}
\label{sec:end}

We have conducted 31 kinetic simulations of six halos (four idealized NFW + two cosmological) across a wide parameter space using \codename. For velocity-independent cross sections and idealized NFW initial conditions, we identify a fitting formula---\cref{eq:tcolfit}---which accurately determines the collapse timescale based on the Knudsen number and relaxation time at the scale radius. (In our simulations, the collapse timescale is defined as the time at which the central density exceeds $25 \times$ its minimum during the collapse.) This generalizes a previous expression in the literature valid only in the LMFP limit. 

Although the collapse timescales as defined can be accurately predicted by this formula, the structure of the collapsing halos and the detailed time evolution of the central density exhibit non-trivial dependencies on Kn. In particular, in the SMFP, the central density grows quasi-exponentially while in the LMFP it diverges super-exponentially (see \cref{fig:col_traj}). 

We have also simulated velocity-dependent, fully differential, and cosmological halos. The general trend of \cref{eq:tcolfit} still holds, but these complications of velocity- and angular-dependence, and cosmological accretion introduce scatter at the factor of few level in the collapse times. For idealized NFW velocity-dependent halos, this scatter can be reduced by taking into account the evolution of the effective conductivity along the collapse trajectory, which we do in \cref{eq:tcolpsi}. The comparison between the collapse timescale across all our simulations with the fitting formulae is shown in \cref{fig:tcol_tdyn_all}. Still, further work is needed to study the cosmological evolution of SIDM in the presence of accretion, especially in the intermediate and SMFP regimes. Specifically, our cosmological simulations (\cref{sec:comsosim}; \cref{tab:gadget_parameters}) explored the extreme limits of collisionless ($N$-body) and hydrodynamical (SPH) evolution. However, neither setup included heat transport via collisions. It would be interesting in the future to run cosmological hydrodynamical simulations with thermal conductivity (which could be calibrated to kinetic simulations). 

These kinetic ({\codename}) simulations are complementary to hydrostatic \enquote{conducting fluid} methods, which are much less computationally expensive than {\codename} in the deep collapse. We have developed a new conducting fluid code called \textsc{SIDMStars} (\cref{sec:fluid}), which solves the same structure equations as previous publicly available tools (e.\,g.~\cite{Gad-Nasr:2023gvf}) using a different numerical approach. 
Specifically, we use entropy rather than internal energy as the state variable. The code alternates between explicit entropy updates at fixed mechanical (hydrostatic) structure, sub-cycled on the CFL time step (\cref{eq:dt_cfl}), and hydrostatic updates at fixed entropy profile $s(m)$, solved by shooting from both boundaries to an intermediate fitting point.

We calibrate \textsc{SIDMStars} to our kinetic {\codename} simulations. While previous fluid codes involve one free parameter $C$ which enters the LMFP conductivity, we find that the inverse sum (\cref{eq:kform_std}) used to interpolate between the LMFP and SMFP conductivity underestimates the collapse time in the intermediate MFP regime. Thus, we calibrate a new interpolation formula (\cref{eq:kform}) with two additional (or three total) free parameters to connect these regimes, which matches the collapse timescales of kinetic simulations better.  

In \textsc{SIDMStars}, a convective zone ($\difs{s}{r} < 0$) can eventually develop outside the collapsing core for constant (velocity-independent and isotropic) cross section runs. We have implemented adiabatic convection in \textsc{SIDMStars} according to a standard prescription in the stellar structure literature \cite{Stahler1980b}. However, the kinetic simulations do not show a flat or inverted entropy gradient for the same parameter values. The immediate relevance of this result is unclear because we do not find convection in the velocity-dependent cross section runs. This potential phenomenology is worth bearing in mind for future work. It may be particularly relevant for models with resonances or dissipation, where the cooling efficiency can be a strong function of position in the halo. At transition or rarefied Knudsen numbers ($\gtrsim 1$) adiabatic convection may not be a reasonable approximation. On the other hand, it seems inevitable that an inverted entropy profile would be smoothed out on a buoyancy timescale.

\subsection*{Upcoming Work}

We close by mentioning two immediate applications of the results presented here. Recently, a joint analysis of 29 quadruply imaged quasars from JWST data found a mild preference for SIDM over CDM. In their analysis, the inferred central density of subhalos is challenging to explain in CDM but can be explained by a population of core-collapsed SIDM halos \cite{gilman2026jwstlensedquasardark}. In a companion to that work \cite{Gilman++26}, the same data are used to infer the dark particle physics parameters under the scenario that the dark matter is self-interacting via a Yukawa interaction carried by a light mediator. That inference is based on \cref{eq:tcolfit}. 

On the other hand, observations of the velocity dispersions in ultra-faint dwarf galaxies are not necessarily consistent with strongly collisional, core-collapsed dark matter halos \cite{2025arXiv250313650A}. In a follow-up paper, we apply the tools developed in this work to study the final state of dark matter halos in that regime and place constraints on the allowed parameter space.

\begin{acknowledgments}
We thank Junwu Huang and Daniel Gilman for useful discussions. Research at the Perimeter Institute is supported in part by the Government of Canada through the Department of Innovation, Science and Economic Development Canada and by the Province of Ontario through the Ministry of Colleges and Universities.
\end{acknowledgments}

\bibliography{sidm}

@CONTROL{apsrev42Control,title="0"}

@article{rackauckas2017differentialequations,
 author = {Rackauckas, Christopher and Nie, Qing},
 doi = {10.5334/jors.151},
 journal = {J. Open Res. Softw.},
 number = {1},
 pages = {15},
 title = "{DifferentialEquations.jl – A Performant and Feature-Rich Ecosystem for Solving Differential Equations in Julia}",
 volume = {5},
 year = {2017},
 publisher={Ubiquity Press}
}

@article{Andrade_2021,
       author = {{Andrade}, Kevin E. and {Fuson}, Jackson and {Gad-Nasr}, Sophia and {Kong}, Demao and {Minor}, Quinn and {Roberts}, M. Grant and {Kaplinghat}, Manoj},
        title = "{A stringent upper limit on dark matter self-interaction cross-section from cluster strong lensing}",
      journal = {\mnras},
         year = 2022,
        month = feb,
       volume = {510},
       number = {1},
        pages = {54-81},
          doi = {10.1093/mnras/stab3241},
archivePrefix = {arXiv},
       eprint = {2012.06611},
 primaryClass = {astro-ph.CO},
       adsurl = {https://ui.adsabs.harvard.edu/abs/2022MNRAS.510...54A}
}

@article{adhikari2022astrophysical,
       author = {{Adhikari}, Susmita and {Banerjee}, Arka and {Boddy}, Kimberly K. and {Cyr-Racine}, Francis-Yan and {Desmond}, Harry and {Dvorkin}, Cora and {Jain}, Bhuvnesh and {Kahlhoefer}, Felix and {Kaplinghat}, Manoj and {Nierenberg}, Anna and {Peter}, Annika H.~G. and {Robertson}, Andrew and {Sakstein}, Jeremy and {Zavala}, Jes{\'u}s},
        title = "{Astrophysical tests of dark matter self-interactions}",
      journal = {Rev. Mod. Phys.},
         year = 2025,
        month = oct,
       volume = {97},
       number = {4},
          eid = {045004},
        pages = {045004},
          doi = {10.1103/m2vm-59y3},
archivePrefix = {arXiv},
       eprint = {2207.10638},
 primaryClass = {astro-ph.CO},
       adsurl = {https://ui.adsabs.harvard.edu/abs/2025RvMP...97d5004A}
}

@article{Nadler_2023,
       author = {{Nadler}, Ethan O. and {Yang}, Daneng and {Yu}, Hai-Bo},
        title = "{A Self-interacting Dark Matter Solution to the Extreme Diversity of Low-mass Halo Properties}",
      journal = {\apjl},
         year = 2023,
        month = dec,
       volume = {958},
       number = {2},
          eid = {L39},
        pages = {L39},
          doi = {10.3847/2041-8213/ad0e09},
archivePrefix = {arXiv},
       eprint = {2306.01830},
 primaryClass = {astro-ph.GA},
       adsurl = {https://ui.adsabs.harvard.edu/abs/2023ApJ...958L..39N}
}

@article{Yu_2026,
       author = {{Yu}, Hai-Bo},
        title = "{Core-Collapsed SIDM Halos as the Common Origin of Dense Perturbers in Lenses, Streams, and Satellites}",
      journal = {\prl},
         year = 2026,
        month = apr,
       volume = {136},
       number = {14},
          eid = {141001},
        pages = {141001},
          doi = {10.1103/txxx-97ln},
archivePrefix = {arXiv},
       eprint = {2510.11006},
 primaryClass = {astro-ph.GA},
       adsurl = {https://ui.adsabs.harvard.edu/abs/2026PhRvL.136n1001Y}
}

@article{Grant25,
       author = {{Roberts}, M. Grant and {Kaplinghat}, Manoj and {Valli}, Mauro and {Yu}, Hai-Bo},
        title = "{Gravothermal collapse and the diversity of galactic rotation curves}",
      journal = {\prd},
         year = 2025,
        month = may,
       volume = {111},
       number = {10},
          eid = {103041},
        pages = {103041},
          doi = {10.1103/PhysRevD.111.103041},
archivePrefix = {arXiv},
       eprint = {2407.15005},
 primaryClass = {astro-ph.GA},
       adsurl = {https://ui.adsabs.harvard.edu/abs/2025PhRvD.111j3041R}
}

@article{Arkani_Hamed_2009,
       author = {{Arkani-Hamed}, Nima and {Finkbeiner}, Douglas P. and {Slatyer}, Tracy R. and {Weiner}, Neal},
        title = "{A theory of dark matter}",
      journal = {\prd},
         year = 2009,
        month = jan,
       volume = {79},
       number = {1},
          eid = {015014},
        pages = {015014},
          doi = {10.1103/PhysRevD.79.015014},
archivePrefix = {arXiv},
       eprint = {0810.0713},
 primaryClass = {hep-ph},
       adsurl = {https://ui.adsabs.harvard.edu/abs/2009PhRvD..79a5014A}
}

@article{Feng_2010,
       author = {{Feng}, Jonathan L. and {Kaplinghat}, Manoj and {Yu}, Hai-Bo},
        title = "{Halo-Shape and Relic-Density Exclusions of Sommerfeld-Enhanced Dark Matter Explanations of Cosmic Ray Excesses}",
      journal = {\prl},
         year = 2010,
        month = apr,
       volume = {104},
       number = {15},
          eid = {151301},
        pages = {151301},
          doi = {10.1103/PhysRevLett.104.151301},
archivePrefix = {arXiv},
       eprint = {0911.0422},
 primaryClass = {hep-ph},
       adsurl = {https://ui.adsabs.harvard.edu/abs/2010PhRvL.104o1301F}
}

@article{Buckley_2010,
       author = {{Buckley}, Matthew R. and {Fox}, Patrick J.},
        title = "{Dark matter self-interactions and light force carriers}",
      journal = {\prd},
         year = 2010,
        month = apr,
       volume = {81},
       number = {8},
          eid = {083522},
        pages = {083522},
          doi = {10.1103/PhysRevD.81.083522},
archivePrefix = {arXiv},
       eprint = {0911.3898},
 primaryClass = {hep-ph},
       adsurl = {https://ui.adsabs.harvard.edu/abs/2010PhRvD..81h3522B}
}

@article{Yang_2023a,
       author = {{Yang}, Daneng and {Nadler}, Ethan O. and {Yu}, Hai-Bo},
        title = "{Strong Dark Matter Self-interactions Diversify Halo Populations within and surrounding the Milky Way}",
      journal = {\apj},
         year = 2023,
        month = jun,
       volume = {949},
       number = {2},
          eid = {67},
        pages = {67},
          doi = {10.3847/1538-4357/acc73e},
archivePrefix = {arXiv},
       eprint = {2211.13768},
 primaryClass = {astro-ph.GA},
       adsurl = {https://ui.adsabs.harvard.edu/abs/2023ApJ...949...67Y}
}

@article{Tulin_2018,
       author = {{Tulin}, Sean and {Yu}, Hai-Bo},
        title = "{Dark matter self-interactions and small scale structure}",
      journal = {\physrep},
         year = 2018,
        month = feb,
       volume = {730},
        pages = {1-57},
          doi = {10.1016/j.physrep.2017.11.004},
archivePrefix = {arXiv},
       eprint = {1705.02358},
 primaryClass = {hep-ph},
       adsurl = {https://ui.adsabs.harvard.edu/abs/2018PhR...730....1T}
}

@article{Robertson_2016,
       author = {{Robertson}, Andrew and {Massey}, Richard and {Eke}, Vincent},
        title = "{What does the Bullet Cluster tell us about self-interacting dark matter?}",
      journal = {\mnras},
         year = 2017,
        month = feb,
       volume = {465},
       number = {1},
        pages = {569-587},
          doi = {10.1093/mnras/stw2670},
archivePrefix = {arXiv},
       eprint = {1605.04307},
 primaryClass = {astro-ph.CO},
       adsurl = {https://ui.adsabs.harvard.edu/abs/2017MNRAS.465..569R}
}

@article{Bento_2000,
       author = {{Bento}, M.~C. and {Bertolami}, O. and {Rosenfeld}, R. and {Teodoro}, L.},
        title = "{Self-interacting dark matter and the Higgs boson}",
      journal = {\prd},
         year = 2000,
        month = aug,
       volume = {62},
       number = {4},
          eid = {041302},
        pages = {041302},
          doi = {10.1103/PhysRevD.62.041302},
archivePrefix = {arXiv},
       eprint = {astro-ph/0003350},
 primaryClass = {astro-ph},
       adsurl = {https://ui.adsabs.harvard.edu/abs/2000PhRvD..62d1302B}
}

@article{Tulin_2013,
       author = {{Tulin}, Sean and {Yu}, Hai-Bo and {Zurek}, Kathryn M.},
        title = "{Beyond collisionless dark matter: Particle physics dynamics for dark matter halo structure}",
      journal = {\prd},
         year = 2013,
        month = jun,
       volume = {87},
       number = {11},
          eid = {115007},
        pages = {115007},
          doi = {10.1103/PhysRevD.87.115007},
archivePrefix = {arXiv},
       eprint = {1302.3898},
 primaryClass = {hep-ph},
       adsurl = {https://ui.adsabs.harvard.edu/abs/2013PhRvD..87k5007T}
}

@article{Loeb_2011,
       author = {{Loeb}, Abraham and {Weiner}, Neal},
        title = "{Cores in Dwarf Galaxies from Dark Matter with a Yukawa Potential}",
      journal = {\prl},
         year = 2011,
        month = apr,
       volume = {106},
       number = {17},
          eid = {171302},
        pages = {171302},
          doi = {10.1103/PhysRevLett.106.171302},
archivePrefix = {arXiv},
       eprint = {1011.6374},
 primaryClass = {astro-ph.CO},
       adsurl = {https://ui.adsabs.harvard.edu/abs/2011PhRvL.106q1302L}
}

@article{Goodman_2000,
       author = {{Goodman}, Jeremy},
        title = "{Repulsive dark matter}",
      journal = {\na},
         year = 2000,
        month = apr,
       volume = {5},
       number = {2},
        pages = {103-107},
          doi = {10.1016/S1384-1076(00)00015-4},
archivePrefix = {arXiv},
       eprint = {astro-ph/0003018},
 primaryClass = {astro-ph},
       adsurl = {https://ui.adsabs.harvard.edu/abs/2000NewA....5..103G}
}

@article{Peebles_Vilenkin_1999,
       author = {{Peebles}, P.~J.~E. and {Vilenkin}, A.},
        title = "{Noninteracting dark matter}",
      journal = {\prd},
         year = 1999,
        month = nov,
       volume = {60},
       number = {10},
          eid = {103506},
        pages = {103506},
          doi = {10.1103/PhysRevD.60.103506},
archivePrefix = {arXiv},
       eprint = {astro-ph/9904396},
 primaryClass = {astro-ph},
       adsurl = {https://ui.adsabs.harvard.edu/abs/1999PhRvD..60j3506P}
}

@misc{Ramos:2025lvk,
    author = "Ramos, Maria and Cohen, Timothy and Lisanti, Mariangela",
    title = "{First-Principles Formalism for Simulating Self-Interacting Dark Matter}",
    eprint = "2512.17998",
    archivePrefix = "arXiv",
    primaryClass = "astro-ph.GA",
    reportNumber = "CERN-TH-2025-263",
    month = "12",
    year = "2025"
}

@misc{Silverman:2026jvu,
    author = "Silverman, Maya and others",
    title = "{Mergers Matter: Gravothermal Collapse in Dwarf Halos with Self-Interacting Dark Matter}",
    eprint = "2606.02566",
    archivePrefix = "arXiv",
    primaryClass = "astro-ph.GA",
    reportNumber = "FERMILAB-PUB-26-0348-T",
    month = "6",
    year = "2026"
}

@ARTICLE{kamionkowski2025numericalevolutionselfgravitatinghalos,
       author = {{Kamionkowski}, Marc and {Sigurdson}, Kris and {Slone}, Oren},
        title = "{Numerical Evolution of Self-Gravitating Halos of Self-Interacting Dark Matter}",
      journal = {\prl},
         year = 2026,
        month = may,
       volume = {136},
       number = {20},
          eid = {201001},
        pages = {201001},
          doi = {10.1103/qht7-5tz3},
archivePrefix = {arXiv},
       eprint = {2506.04334},
 primaryClass = {astro-ph.CO},
       adsurl = {https://ui.adsabs.harvard.edu/abs/2026PhRvL.136t1001K}
}

@misc{pal2024nonlinearsolve,
       author = {{Pal}, Avik and {Holtorf}, Flemming and {Larsson}, Axel and {Loman}, Torkel and {Utkarsh} and {Sch{\"a}efer}, Frank and {Qu}, Qingyu and {Edelman}, Alan and {Rackauckas}, Chris},
        title = "{NonlinearSolve.jl: High-Performance and Robust Solvers for Systems of Nonlinear Equations in Julia}",
      journal = {arXiv e-prints},
         year = 2024,
        month = mar,
          eid = {arXiv:2403.16341},
        pages = {arXiv:2403.16341},
archivePrefix = {arXiv},
       eprint = {2403.16341},
 primaryClass = {cs.NA},
       adsurl = {https://ui.adsabs.harvard.edu/abs/2024arXiv240316341P}
}

@misc{Meng:2026hzv,
       author = {{Meng}, Zhe and {Chen}, Tan and {Zhu}, Bocheng and {Zhou}, Fan and {Hu}, Bin and {Gao}, Liang and {Cai}, Rong-Gen},
        title = "{Spherically Symmetric Fluid Simulations of Black Hole Accretion in Self-Interacting Dark Matter Halos}",
      journal = {arXiv e-prints},
         year = 2026,
        month = jul,
          eid = {arXiv:2607.02151},
        pages = {arXiv:2607.02151},
archivePrefix = {arXiv},
       eprint = {2607.02151},
 primaryClass = {astro-ph.CO},
       adsurl = {https://ui.adsabs.harvard.edu/abs/2026arXiv260702151M}
}

@misc{zeng2026feedingvoidcoevolutionsidmseeded,
       author = {{Carton Zeng}, Zhichao and {Peter}, Annika H.~G. and {Strigari}, Louis E. and {Benson}, Andrew and {Du}, Xiaolong and {Mace}, Charlie},
        title = "{Feeding the Void: Co-evolution of the SIDM-seeded Black Hole and the dark halo after core-collapse}",
      journal = {arXiv e-prints},
         year = 2026,
        month = sep,
          eid = {arXiv:2609.22758},
        pages = {arXiv:2609.22758},
archivePrefix = {arXiv},
       eprint = {2609.22758},
 primaryClass = {astro-ph.CO},
       adsurl = {https://ui.adsabs.harvard.edu/abs/2026arXiv260922758C}
}

@article{Gu:2026zzq,
    author = "Gu, Hua-Peng and Jiang, Fangzhou and Chen, Xian and Li, Ran",
    title = "{Nonequilibrium relativistic core collapse of self-interacting dark matter halos: Limits on the seed black hole mass}",
    eprint = "2601.17117",
    archivePrefix = "arXiv",
    primaryClass = "astro-ph.CO",
    doi = "10.1103/6sl3-kjzf",
    journal = "Phys. Rev. D",
    volume = "113",
    number = "10",
    pages = "103038",
    year = "2026"
}

@article{Elbert_2015,
       author = {{Elbert}, Oliver D. and {Bullock}, James S. and {Garrison-Kimmel}, Shea and {Rocha}, Miguel and {O{\~n}orbe}, Jose and {Peter}, Annika H.~G.},
        title = "{Core formation in dwarf haloes with self-interacting dark matter: no fine-tuning necessary}",
      journal = {\mnras},
         year = 2015,
        month = oct,
       volume = {453},
       number = {1},
        pages = {29-37},
          doi = {10.1093/mnras/stv1470},
archivePrefix = {arXiv},
       eprint = {1412.1477},
 primaryClass = {astro-ph.GA},
       adsurl = {https://ui.adsabs.harvard.edu/abs/2015MNRAS.453...29E}
}

@article{Nishikawa_2020,
       author = {{Nishikawa}, Hiroya and {Boddy}, Kimberly K. and {Kaplinghat}, Manoj},
        title = "{Accelerated core collapse in tidally stripped self-interacting dark matter halos}",
      journal = {\prd},
         year = 2020,
        month = mar,
       volume = {101},
       number = {6},
          eid = {063009},
        pages = {063009},
          doi = {10.1103/PhysRevD.101.063009},
archivePrefix = {arXiv},
       eprint = {1901.00499},
 primaryClass = {astro-ph.GA},
       adsurl = {https://ui.adsabs.harvard.edu/abs/2020PhRvD.101f3009N}
}

@article{Garrison_Kimmel_2013,
       author = {{Garrison-Kimmel}, Shea and {Rocha}, Miguel and {Boylan-Kolchin}, Michael and {Bullock}, James S. and {Lally}, Jaspreet},
        title = "{Can feedback solve the too-big-to-fail problem?}",
      journal = {\mnras},
         year = 2013,
        month = aug,
       volume = {433},
       number = {4},
        pages = {3539-3546},
          doi = {10.1093/mnras/stt984},
archivePrefix = {arXiv},
       eprint = {1301.3137},
 primaryClass = {astro-ph.CO},
       adsurl = {https://ui.adsabs.harvard.edu/abs/2013MNRAS.433.3539G}
}

@ARTICLE{Lynden-Bell80,
       author = {{Lynden-Bell}, D. and {Eggleton}, P.~P.},
        title = "{On the consequences of the gravothermal catastrophe}",
      journal = {\mnras},
         year = 1980,
        month = may,
       volume = {191},
        pages = {483-498},
          doi = {10.1093/mnras/191.3.483},
       adsurl = {https://ui.adsabs.harvard.edu/abs/1980MNRAS.191..483L}
}

@ARTICLE{Pollack2015,
       author = {{Pollack}, Jason and {Spergel}, David N. and {Steinhardt}, Paul J.},
        title = "{Supermassive Black Holes from Ultra-strongly Self-interacting Dark Matter}",
      journal = {\apj},
         year = 2015,
        month = may,
       volume = {804},
       number = {2},
          eid = {131},
        pages = {131},
          doi = {10.1088/0004-637X/804/2/131},
archivePrefix = {arXiv},
       eprint = {1501.00017},
 primaryClass = {astro-ph.CO},
       adsurl = {https://ui.adsabs.harvard.edu/abs/2015ApJ...804..131P}
}

@ARTICLE{palubski2024numericalchallengesmodelinggravothermal,
       author = {{Palubski}, Igor and {Slone}, Oren and {Kaplinghat}, Manoj and {Lisanti}, Mariangela and {Jiang}, Fangzhou},
        title = "{Numerical challenges in modeling gravothermal collapse in Self-Interacting Dark Matter halos}",
      journal = {\jcap},
         year = 2024,
        month = sep,
       volume = {2024},
       number = {9},
          eid = {074},
        pages = {074},
          doi = {10.1088/1475-7516/2024/09/074},
archivePrefix = {arXiv},
       eprint = {2402.12452},
 primaryClass = {astro-ph.CO},
       adsurl = {https://ui.adsabs.harvard.edu/abs/2024JCAP...09..074P}
}

@ARTICLE{Kamada2017,
       author = {{Kamada}, Ayuki and {Kaplinghat}, Manoj and {Pace}, Andrew B. and {Yu}, Hai-Bo},
        title = "{Self-Interacting Dark Matter Can Explain Diverse Galactic Rotation Curves}",
      journal = {\prl},
         year = 2017,
        month = sep,
       volume = {119},
       number = {11},
          eid = {111102},
        pages = {111102},
          doi = {10.1103/PhysRevLett.119.111102},
archivePrefix = {arXiv},
       eprint = {1611.02716},
 primaryClass = {astro-ph.GA},
       adsurl = {https://ui.adsabs.harvard.edu/abs/2017PhRvL.119k1102K}
}

@article{Peebles_2000,
       author = {{Peebles}, P.~J.~E.},
        title = "{Fluid Dark Matter}",
      journal = {\apjl},
         year = 2000,
        month = may,
       volume = {534},
       number = {2},
        pages = {L127-L129},
          doi = {10.1086/312677},
archivePrefix = {arXiv},
       eprint = {astro-ph/0002495},
 primaryClass = {astro-ph},
       adsurl = {https://ui.adsabs.harvard.edu/abs/2000ApJ...534L.127P}
}

@article{Balberg_2002a,
       author = {{Balberg}, Shmuel and {Shapiro}, Stuart L. and {Inagaki}, Shogo},
        title = "{Self-Interacting Dark Matter Halos and the Gravothermal Catastrophe}",
      journal = {\apj},
         year = 2002,
        month = apr,
       volume = {568},
       number = {2},
        pages = {475-487},
          doi = {10.1086/339038},
archivePrefix = {arXiv},
       eprint = {astro-ph/0110561},
 primaryClass = {astro-ph},
       adsurl = {https://ui.adsabs.harvard.edu/abs/2002ApJ...568..475B}
}

@article{Yang_2022,
       author = {{Yang}, Daneng and {Yu}, Hai-Bo},
        title = "{Gravothermal evolution of dark matter halos with differential elastic scattering}",
      journal = {\jcap},
         year = 2022,
        month = sep,
       volume = {2022},
       number = {9},
          eid = {077},
        pages = {077},
          doi = {10.1088/1475-7516/2022/09/077},
archivePrefix = {arXiv},
       eprint = {2205.03392},
 primaryClass = {astro-ph.CO},
       adsurl = {https://ui.adsabs.harvard.edu/abs/2022JCAP...09..077Y}
}

@ARTICLE{fischer2025accuratelysimulatingcorecollapseselfinteracting,
       author = {{Fischer}, Moritz S. and {Yu}, Hai-Bo and {Dolag}, Klaus},
        title = "{Accurately simulating core-collapse self-interacting dark matter halos}",
      journal = {\aap},
         year = 2025,
        month = nov,
       volume = {703},
          eid = {A234},
        pages = {A234},
          doi = {10.1051/0004-6361/202556189},
archivePrefix = {arXiv},
       eprint = {2506.06269},
 primaryClass = {astro-ph.CO},
       adsurl = {https://ui.adsabs.harvard.edu/abs/2025A&A...703A.234F}
}

@article{Essig_2019,
       author = {{Essig}, Rouven and {McDermott}, Samuel D. and {Yu}, Hai-Bo and {Zhong}, Yi-Ming},
        title = "{Constraining Dissipative Dark Matter Self-Interactions}",
      journal = {\prl},
         year = 2019,
        month = sep,
       volume = {123},
       number = {12},
          eid = {121102},
        pages = {121102},
          doi = {10.1103/PhysRevLett.123.121102},
archivePrefix = {arXiv},
       eprint = {1809.01144},
 primaryClass = {hep-ph},
       adsurl = {https://ui.adsabs.harvard.edu/abs/2019PhRvL.123l1102E}
}

@article{Balberg_2002,
       author = {{Balberg}, Shmuel and {Shapiro}, Stuart L.},
        title = "{Gravothermal Collapse of Self-Interacting Dark Matter Halos and the Origin of Massive Black Holes}",
      journal = {\prl},
         year = 2002,
        month = mar,
       volume = {88},
       number = {10},
          eid = {101301},
        pages = {101301},
          doi = {10.1103/PhysRevLett.88.101301},
archivePrefix = {arXiv},
       eprint = {astro-ph/0111176},
 primaryClass = {astro-ph},
       adsurl = {https://ui.adsabs.harvard.edu/abs/2002PhRvL..88j1301B}
}

@ARTICLE{mace2025calibratingsidmgravothermalcatastrophe,
       author = {{Mace}, Charlie and {Yang}, Shengqi and {Zeng}, Zhichao Carton and {Peter}, Annika H.~G. and {Du}, Xiaolong and {Benson}, Andrew},
        title = "{Calibrating the self-interacting dark matter gravothermal catastrophe with $N$-body simulations}",
      journal = {\prd},
         year = 2026,
        month = may,
       volume = {113},
       number = {10},
          eid = {103011},
        pages = {103011},
          doi = {10.1103/x3lh-xfx9},
archivePrefix = {arXiv},
       eprint = {2504.13004},
 primaryClass = {astro-ph.GA},
       adsurl = {https://ui.adsabs.harvard.edu/abs/2026PhRvD.113j3011M}
}

@article{Yang_2023,
       author = {{Yang}, Shengqi and {Du}, Xiaolong and {Zeng}, Zhichao Carton and {Benson}, Andrew and {Jiang}, Fangzhou and {Nadler}, Ethan O. and {Peter}, Annika H.~G.},
        title = "{Gravothermal Solutions of SIDM Halos: Mapping from Constant to Velocity-dependent Cross Section}",
      journal = {\apj},
         year = 2023,
        month = mar,
       volume = {946},
       number = {1},
          eid = {47},
        pages = {47},
          doi = {10.3847/1538-4357/acbd49},
archivePrefix = {arXiv},
       eprint = {2205.02957},
 primaryClass = {astro-ph.CO},
       adsurl = {https://ui.adsabs.harvard.edu/abs/2023ApJ...946...47Y}
}

@ARTICLE{mace2024convergencetestsselfinteractingdark,
       author = {{Mace}, Charlie and {Zeng}, Zhichao Carton and {Peter}, Annika H.~G. and {Du}, Xiaolong and {Yang}, Shengqi and {Benson}, Andrew and {Vogelsberger}, Mark},
        title = "{Convergence tests of self-interacting dark matter simulations}",
      journal = {\prd},
         year = 2024,
        month = dec,
       volume = {110},
       number = {12},
          eid = {123024},
        pages = {123024},
          doi = {10.1103/PhysRevD.110.123024},
archivePrefix = {arXiv},
       eprint = {2402.01604},
 primaryClass = {astro-ph.GA},
       adsurl = {https://ui.adsabs.harvard.edu/abs/2024PhRvD.110l3024M}
}

@ARTICLE{Stahler1980b,
       author = {{Stahler}, S.~W. and {Shu}, F.~H. and {Taam}, R.~E.},
        title = "{The evolution of protostars. II – The hydrostatic core}",
      journal = {\apj},
         year = 1980,
        month = nov,
       volume = {242},
        pages = {226-241},
          doi = {10.1086/158459},
       adsurl = {https://ui.adsabs.harvard.edu/abs/1980ApJ...242..226S}
}

@ARTICLE{Hosokawa2009,
       author = {{Hosokawa}, Takashi and {Omukai}, Kazuyuki},
        title = "{Evolution of Massive Protostars with High Accretion Rates}",
      journal = {\apj},
         year = 2009,
        month = jan,
       volume = {691},
       number = {1},
        pages = {823-846},
          doi = {10.1088/0004-637X/691/1/823},
archivePrefix = {arXiv},
       eprint = {0806.4122},
 primaryClass = {astro-ph},
       adsurl = {https://ui.adsabs.harvard.edu/abs/2009ApJ...691..823H}
}

@article{Gad-Nasr:2023gvf,
    author = "Gad-Nasr, Sophia and Boddy, Kimberly K. and Kaplinghat, Manoj and Outmezguine, Nadav Joseph and Sagunski, Laura",
    title = "{On the late-time evolution of velocity-dependent self-interacting dark matter halos}",
    eprint = "2312.09296",
    archivePrefix = "arXiv",
    primaryClass = "astro-ph.GA",
    doi = "10.1088/1475-7516/2024/05/131",
    journal = "\jcap",
    volume = "05",
    pages = "131",
    year = "2024"
}

@ARTICLE{Gilman++26,
	author = {{Gilman}, Daniel and others},
	title = "{Hints of self-interacting dark matter from strong gravitational lensing}",
	journal = "to appear on arXiv",
	year = 2026,
	primaryClass = {astro-ph.CO}
}

@book{bird1994molecular,
    author = {Bird, G A},
    title = {Molecular Gas Dynamics and the Direct Simulation of Gas Flows},
    publisher = {Oxford University Press},
    year = {1994},
    month = {05},
    isbn = {9780198561958},
    doi = {10.1093/oso/9780198561958.001.0001}
}

@BOOK{Chapman70,
    isbn = {9780521408448},
       author = {{Chapman}, Sydeny and {Cowling}, T.~G.},
        title = "{The Mathematical Theory of Non-uniform Gases}",
    subtitle = "{An Account of the Kinetic Theory of Viscosity, Thermal Conduction and Diffusion in Gases}",
         year = 1970,
       adsurl = {https://ui.adsabs.harvard.edu/abs/1970mtnu.book.....C},
    publisher= {Cambridge University Press}
}

@article{Outmezguine:2022bhq,
    author = "Outmezguine, Nadav Joseph and Boddy, Kimberly K. and Gad-Nasr, Sophia and Kaplinghat, Manoj and Sagunski, Laura",
    title = "{Universal gravothermal evolution of isolated self-interacting dark matter halos for velocity-dependent cross-sections}",
    eprint = "2204.06568",
    archivePrefix = "arXiv",
    primaryClass = "astro-ph.GA",
    doi = "10.1093/mnras/stad1705",
    journal = "\mnras",
    volume = "523",
    number = "3",
    pages = "4786--4800",
    year = "2023"
}

@article{Bullock_2017,
       author = {{Bullock}, James S. and {Boylan-Kolchin}, Michael},
        title = "{Small-Scale Challenges to the {\ensuremath{\Lambda}}CDM Paradigm}",
      journal = {\araa},
         year = 2017,
        month = aug,
       volume = {55},
       number = {1},
        pages = {343-387},
          doi = {10.1146/annurev-astro-091916-055313},
archivePrefix = {arXiv},
       eprint = {1707.04256},
 primaryClass = {astro-ph.CO},
       adsurl = {https://ui.adsabs.harvard.edu/abs/2017ARA&A..55..343B}
}

@article{Fry_2015,
       author = {{Fry}, A. Bastidas and {Governato}, F. and {Pontzen}, A. and {Quinn}, T. and {Tremmel}, M. and {Anderson}, L. and {Menon}, H. and {Brooks}, A.~M. and {Wadsley}, J.},
        title = "{All about baryons: revisiting SIDM predictions at small halo masses}",
      journal = {\mnras},
         year = 2015,
        month = sep,
       volume = {452},
       number = {2},
        pages = {1468-1479},
          doi = {10.1093/mnras/stv1330},
archivePrefix = {arXiv},
       eprint = {1501.00497},
 primaryClass = {astro-ph.CO},
       adsurl = {https://ui.adsabs.harvard.edu/abs/2015MNRAS.452.1468F}
}

@article{Vogelsberger_2012,
       author = {{Vogelsberger}, Mark and {Zavala}, Jesus and {Loeb}, Abraham},
        title = "{Subhaloes in self-interacting galactic dark matter haloes}",
      journal = {\mnras},
         year = 2012,
        month = jul,
       volume = {423},
       number = {4},
        pages = {3740-3752},
          doi = {10.1111/j.1365-2966.2012.21182.x10.1002/asna.19141991009},
archivePrefix = {arXiv},
       eprint = {1201.5892},
 primaryClass = {astro-ph.CO},
       adsurl = {https://ui.adsabs.harvard.edu/abs/2012MNRAS.423.3740V}
}

@article{Vogelsberger_2014,
       author = {{Vogelsberger}, Mark and {Zavala}, Jesus and {Simpson}, Christine and {Jenkins}, Adrian},
        title = "{Dwarf galaxies in CDM and SIDM with baryons: observational probes of the nature of dark matter}",
      journal = {\mnras},
         year = 2014,
        month = nov,
       volume = {444},
       number = {4},
        pages = {3684-3698},
          doi = {10.1093/mnras/stu1713},
archivePrefix = {arXiv},
       eprint = {1405.5216},
 primaryClass = {astro-ph.CO},
       adsurl = {https://ui.adsabs.harvard.edu/abs/2014MNRAS.444.3684V}
}

@ARTICLE{markevitch_direct_2004,
       author = {{Markevitch}, M. and {Gonzalez}, A.~H. and {Clowe}, D. and {Vikhlinin}, A. and {Forman}, W. and {Jones}, C. and {Murray}, S. and {Tucker}, W.},
        title = "{Direct Constraints on the Dark Matter Self-Interaction Cross Section from the Merging Galaxy Cluster 1E 0657-56}",
      journal = {\apj},
         year = 2004,
        month = may,
       volume = {606},
       number = {2},
        pages = {819-824},
          doi = {10.1086/383178},
archivePrefix = {arXiv},
       eprint = {astro-ph/0309303},
 primaryClass = {astro-ph},
       adsurl = {https://ui.adsabs.harvard.edu/abs/2004ApJ...606..819M}
}

@misc{gilman2026jwstlensedquasardark,
       author = {{Gilman}, D. and {Nierenberg}, A.~M. and {Gurian}, J. and {Paugnat}, H. and {Gannon}, C. and {Martinez}, M.~N. and {Treu}, T. and {Abazajian}, K.~N. and {Anguita}, T. and {Bennert}, V.~N. and {Benson}, A.~J. and {Birrer}, S. and {Djorgovski}, S.~G. and {Hoenig}, S.~F. and {Keeley}, R.~E. and {Kusenko}, A. and {Millon}, M. and {Morishita}, T. and {Moustakas}, L.~A. and {Mozumdar}, P. and {Paris}, D. and {Sheu}, W. and {Sluse}, D. and {Wong}, K.~C.},
        title = "{JWST lensed quasar dark matter survey V: Hints of self-interacting dark matter from 29 quadruply imaged quasars}",
      journal = {arXiv e-prints},
         year = 2026,
        month = sep,
          eid = {arXiv:2609.35974},
        pages = {arXiv:2609.35974},
archivePrefix = {arXiv},
       eprint = {2609.35974},
 primaryClass = {astro-ph.CO},
       adsurl = {https://ui.adsabs.harvard.edu/abs/2026arXiv260935974G}
}

@ARTICLE{NFW,
       author = {{Navarro}, Julio F. and {Frenk}, Carlos S. and {White}, Simon D.~M.},
        title = "{A Universal Density Profile from Hierarchical Clustering}",
      journal = {\apj},
         year = 1997,
        month = dec,
       volume = {490},
       number = {2},
        pages = {493-508},
          doi = {10.1086/304888},
archivePrefix = {arXiv},
       eprint = {astro-ph/9611107},
 primaryClass = {astro-ph},
       adsurl = {https://ui.adsabs.harvard.edu/abs/1997ApJ...490..493N}
}

@article{Springel_2021,
       author = {{Springel}, Volker and {Pakmor}, R{\"u}diger and {Zier}, Oliver and {Reinecke}, Martin},
        title = "{Simulating cosmic structure formation with the GADGET-4 code}",
      journal = {\mnras},
         year = 2021,
        month = sep,
       volume = {506},
       number = {2},
        pages = {2871-2949},
          doi = {10.1093/mnras/stab1855},
archivePrefix = {arXiv},
       eprint = {2010.03567},
 primaryClass = {astro-ph.IM},
       adsurl = {https://ui.adsabs.harvard.edu/abs/2021MNRAS.506.2871S}
}

@misc{ngenic_2015,
       author = {{Springel}, Volker},
        title = "{N-GenIC: Cosmological structure initial conditions}",
 howpublished = {Astrophysics Source Code Library, record ascl:1502.003},
         year = 2015,
        month = feb,
          eid = {ascl:1502.003},
       adsurl = {https://ui.adsabs.harvard.edu/abs/2015ascl.soft02003S},
    url = {https://ascl.net/1502.003}
}

@ARTICLE{Henyey_1964,
       author = {{Henyey}, L.~G. and {Forbes}, J.~E. and {Gould}, N.~L.},
        title = "{A New Method of Automatic Computation of Stellar Evolution.}",
      journal = {\apj},
         year = 1964,
        month = jan,
       volume = {139},
        pages = {306},
          doi = {10.1086/147754},
       adsurl = {https://ui.adsabs.harvard.edu/abs/1964ApJ...139..306H}
}

@ARTICLE{Planck_2018,
    collaboration = {Planck},
       author = {{Aghanim}, N. and {Akrami}, Y. and {Ashdown}, M. and {Aumont}, J. and {Baccigalupi}, C. and {Ballardini}, M. and {Banday}, A.~J. and {Barreiro}, R.~B. and {Bartolo}, N. and {Basak}, S. and {Battye}, R. and {Benabed}, K. and {Bernard}, J.-P. and {Bersanelli}, M. and {Bielewicz}, P. and {Bock}, J.~J. and {Bond}, J.~R. and {Borrill}, J. and {Bouchet}, F.~R. and {Boulanger}, F. and {Bucher}, M. and {Burigana}, C. and {Butler}, R.~C. and {Calabrese}, E. and {Cardoso}, J.-F. and {Carron}, J. and {Challinor}, A. and {Chiang}, H.~C. and {Chluba}, J. and {Colombo}, L.~P.~L. and {Combet}, C. and {Contreras}, D. and {Crill}, B.~P. and {Cuttaia}, F. and {de Bernardis}, P. and {de Zotti}, G. and {Delabrouille}, J. and {Delouis}, J.-M. and {Di Valentino}, E. and {Diego}, J.~M. and {Dor{\'e}}, O. and {Douspis}, M. and {Ducout}, A. and {Dupac}, X. and {Dusini}, S. and {Efstathiou}, G. and {Elsner}, F. and {En{\ss}lin}, T.~A. and {Eriksen}, H.~K. and {Fantaye}, Y. and {Farhang}, M. and {Fergusson}, J. and {Fernandez-Cobos}, R. and {Finelli}, F. and {Forastieri}, F. and {Frailis}, M. and {Fraisse}, A.~A. and {Franceschi}, E. and {Frolov}, A. and {Galeotta}, S. and {Galli}, S. and {Ganga}, K. and {G{\'e}nova-Santos}, R.~T. and {Gerbino}, M. and {Ghosh}, T. and {Gonz{\'a}lez-Nuevo}, J. and {G{\'o}rski}, K.~M. and {Gratton}, S. and {Gruppuso}, A. and {Gudmundsson}, J.~E. and {Hamann}, J. and {Handley}, W. and {Hansen}, F.~K. and {Herranz}, D. and {Hildebrandt}, S.~R. and {Hivon}, E. and {Huang}, Z. and {Jaffe}, A.~H. and {Jones}, W.~C. and {Karakci}, A. and {Keih{\"a}nen}, E. and {Keskitalo}, R. and {Kiiveri}, K. and {Kim}, J. and {Kisner}, T.~S. and {Knox}, L. and {Krachmalnicoff}, N. and {Kunz}, M. and {Kurki-Suonio}, H. and {Lagache}, G. and {Lamarre}, J.-M. and {Lasenby}, A. and {Lattanzi}, M. and {Lawrence}, C.~R. and {Le Jeune}, M. and {Lemos}, P. and {Lesgourgues}, J. and {Levrier}, F. and {Lewis}, A. and {Liguori}, M. and {Lilje}, P.~B. and {Lilley}, M. and {Lindholm}, V. and {L{\'o}pez-Caniego}, M. and {Lubin}, P.~M. and {Ma}, Y.-Z. and {Mac{\'\i}as-P{\'e}rez}, J.~F. and {Maggio}, G. and {Maino}, D. and {Mandolesi}, N. and {Mangilli}, A. and {Marcos-Caballero}, A. and {Maris}, M. and {Martin}, P.~G. and {Martinelli}, M. and {Mart{\'\i}nez-Gonz{\'a}lez}, E. and {Matarrese}, S. and {Mauri}, N. and {McEwen}, J.~D. and {Meinhold}, P.~R. and {Melchiorri}, A. and {Mennella}, A. and {Migliaccio}, M. and {Millea}, M. and {Mitra}, S. and {Miville-Desch{\^e}nes}, M.-A. and {Molinari}, D. and {Montier}, L. and {Morgante}, G. and {Moss}, A. and {Natoli}, P. and {N{\o}rgaard-Nielsen}, H.~U. and {Pagano}, L. and {Paoletti}, D. and {Partridge}, B. and {Patanchon}, G. and {Peiris}, H.~V. and {Perrotta}, F. and {Pettorino}, V. and {Piacentini}, F. and {Polastri}, L. and {Polenta}, G. and {Puget}, J.-L. and {Rachen}, J.~P. and {Reinecke}, M. and {Remazeilles}, M. and {Renzi}, A. and {Rocha}, G. and {Rosset}, C. and {Roudier}, G. and {Rubi{\~n}o-Mart{\'\i}n}, J.~A. and {Ruiz-Granados}, B. and {Salvati}, L. and {Sandri}, M. and {Savelainen}, M. and {Scott}, D. and {Shellard}, E.~P.~S. and {Sirignano}, C. and {Sirri}, G. and {Spencer}, L.~D. and {Sunyaev}, R. and {Suur-Uski}, A.-S. and {Tauber}, J.~A. and {Tavagnacco}, D. and {Tenti}, M. and {Toffolatti}, L. and {Tomasi}, M. and {Trombetti}, T. and {Valenziano}, L. and {Valiviita}, J. and {Van Tent}, B. and {Vibert}, L. and {Vielva}, P. and {Villa}, F. and {Vittorio}, N. and {Wandelt}, B.~D. and {Wehus}, I.~K. and {White}, M. and {White}, S.~D.~M. and {Zacchei}, A. and {Zonca}, A.},
        title = "{Planck 2018 results. VI. Cosmological parameters}",
      journal = {\aap},
         year = 2020,
        month = sep,
       volume = {641},
          eid = {A6},
        pages = {A6},
          doi = {10.1051/0004-6361/201833910},
archivePrefix = {arXiv},
       eprint = {1807.06209},
 primaryClass = {astro-ph.CO},
       adsurl = {https://ui.adsabs.harvard.edu/abs/2020A&A...641A...6P}
}

@article{Gurian_2025,
       author = {{Gurian}, James and {May}, Simon},
        title = "{Core Collapse Beyond the Fluid Approximation: The Late Evolution of Self-Interacting Dark Matter Halos}",
      journal = {\prl},
         year = 2025,
        month = nov,
       volume = {135},
       number = {22},
          eid = {221001},
        pages = {221001},
          doi = {10.1103/2ycz-3fvv},
archivePrefix = {arXiv},
       eprint = {2505.15903},
 primaryClass = {astro-ph.CO},
       adsurl = {https://ui.adsabs.harvard.edu/abs/2025PhRvL.135v1001G}
}

@ARTICLE{1997ApJ...490..493N,
       author = {{Navarro}, Julio F. and {Frenk}, Carlos S. and {White}, Simon D.~M.},
        title = "{A Universal Density Profile from Hierarchical Clustering}",
      journal = {\apj},
         year = 1997,
        month = dec,
       volume = {490},
       number = {2},
        pages = {493-508},
          doi = {10.1086/304888},
archivePrefix = {arXiv},
       eprint = {astro-ph/9611107},
 primaryClass = {astro-ph},
       adsurl = {https://ui.adsabs.harvard.edu/abs/1997ApJ...490..493N}
}

@Book{Kippenhahn2012,
  author    = {Rudolf Kippenhahn and Alfred Weigert and Achim Weiss},
  publisher = {Springer Berlin Heidelberg},
  title     = {Stellar Structure and Evolution},
  year      = {2012},
  doi       = {10.1007/978-3-642-30304-3},
}

@misc{2025arXiv250313650A,
       author = {{Ando}, Shin'ichiro and {Hayashi}, Kohei and {Horigome}, Shunichi and {Ibe}, Masahiro and {Shirai}, Satoshi},
        title = "{Stringent Constraints on Self-Interacting Dark Matter Using Milky-Way Satellite Galaxies Kinematics}",
      journal = {arXiv e-prints},
         year = 2025,
        month = mar,
          eid = {arXiv:2503.13650},
        pages = {arXiv:2503.13650},
archivePrefix = {arXiv},
       eprint = {2503.13650},
 primaryClass = {astro-ph.CO},
       adsurl = {https://ui.adsabs.harvard.edu/abs/2025arXiv250313650A}
}


\appendix

\section{Implementation of Velocity-Dependent and Differential Scattering in {\normalfont\codename}}
\label{app:vdep_imp}

{\codename} is based on the Direct Simulation Monte Carlo (DSMC) approach, which involves partitioning the computational domain into spatial cells and performing stochastic, pairwise scatterings between particles in the same cell. For velocity-dependent, anisotropic cross sections, the algorithm within each cell and at each time step is as follows:
\begin{enumerate}
    \item Compute an upper bound on the scattering rate in the cell:
    \begin{equation}
        M = \left\lceil\frac{1}{2}N \rho\Sigma \Delta t\right\rceil,
    \end{equation}
    where $N$ is the number of particles, $\rho$ is the mass density, $\Sigma = 3 \langle\sigma v\rangle$ (see discussion below) and $\Delta t$ is the time step. The factor $1/2$ accounts for double counting pairs, and $\lceil\cdot\rceil$ is the ceiling function rounding $M$ to the next integer larger than $\frac{1}{2}N \rho\Sigma \Delta t$.
    \item Draw two particles in the cell, $i$ and $j$. Compute $\sigma_{v_{\mathrm{rel}}}/(a\Sigma)$, where $a = M / (\frac{1}{2}N \rho\Sigma \Delta t)$ is the factor scaling $M$ to the next integer, and a uniform random number $c \in [0,1)$. If $c < \sigma_{v_{\mathrm{rel}}}/(a\Sigma)$:
    \begin{itemize} 
        \item Sample a random angle from $\difs{\sigma}{\cos(\theta)}$ (via rejection sampling)
        \item Generate a pair of velocity directions $\vec{v}_i$ and $\vec{v}_j$ with $\hat i \cdot \hat j = \cos(\theta)$ (with uniform random azimuth) and $||\vec{v}_i|| = ||\vec{v}_j|| = v_{\mathrm{rel}}/2$, where $v_{\mathrm{rel}}$ is the relative velocity of the particles. These are the center of momentum velocities of the particles. Transform back to the halo frame. 
        \item Assign new velocities to the particles $i$ and $j$ by transforming $\vec{v}_i$ and $\vec{v}_j$ to the halo frame.
    \end{itemize}
    \item Repeat step 2.\ $M$ times. 
\end{enumerate}
For velocity-dependent but isotropic cross sections, the procedure is the same except that the angle is chosen isotropically. 

Generally, the DSMC algorithm demands the following resolution conditions: 
\begin{itemize}
    \item $ l < \lambda$, where $l$ is the cell size and $\lambda$ is the mean free path,
    \item $\Delta t < 1/(\rho \langle \sigma v\rangle$), and
    \item $\Sigma >\sigma(v_{\mathrm{rel}}) \, v_{\mathrm{rel}}$ for all pairs in the cell.
\end{itemize}

For the $v^{-4}$-type cross sections studied here, these resolution constraints, computed naively, become extremely restrictive. In particular, note that the condition on the cell size requires an increasing number of particles to adequately resolve the distribution function in a given cell as $\sigma$ increases (in our runs, a cell is by default split based on $\lambda$ only if it contains at least 32~particles). However, this cost is dominated by an extremely large number of low relative-velocity scatterings which have only a minor effect on the velocity distribution function. For this reason, we have relaxed these requirement to resolving the effective conductivity cross section, making the substitution $\langle \sigma v\rangle \rightarrow \Gamma_5$ in the spatial and temporal resolution conditions. We have also relaxed the definition of the majorant $M$, adopting an upper bound on the total scattering rate in the cell rather than on the scattering rate for all relative velocity pairs. This means that in practice pairs with $\sigma v_{\mathrm{rel}} >\Sigma$ are always scattered, and the total rate of such scatterings is underestimated. However, because (for $v^{-4}$-type cross sections) these are low relative velocity pairs we expect minor impact on the transport. The calculation of $M$ should be critically re-evaluated if simulating resonant or rising power law cross sections. The numerical choices are summarized in \cref{tab:cond_numerics}.

\begin{table}
    \centering
    \caption{Collision numerics. For $\tilde{\sigma}_{\mathrm{m}}$, the adopted numerics use the $w \ll \nu$ limit $\tilde{\sigma}_{\mathrm{m}} \approx \sigma_0(w/\nu)^4/96$.}
    \smallskip
    \setlength{\tabcolsep}{6pt}
    \begin{tabular}{lll}
    \toprule
    Quantity & Conservative & Adopted
    \\
    & & ($v^{-4}$ collapse runs)
    \\
    \midrule
    $\Sigma$  &
      $\max_{v}\, \sigma(v)\,v$ &
      $3\,\langle\sigma v_{\mathrm{rel}}\rangle$ \\[2pt]
    Time step $\Delta t$ &
      $\dfrac{1}{3}\min_{\mathrm{cells}}\dfrac{1}{n\,\Sigma}$ &
      $\dfrac{1}{3}\min_{\mathrm{cells}}\dfrac{1}{n\,\tilde{\sigma}_{\mathrm{m}}\,\sqrt{3}\nu}$ \\[8pt]
    Cell size $\Delta x$ &
      $\dfrac{1}{n\,\sigma\mleft(\sqrt{2}\,\nu\mright)}$ &
      $\dfrac{1}{n\,\tilde{\sigma}_{\mathrm{m}}(\nu)}$ \\[8pt]
    \bottomrule
    \end{tabular}
    \label{tab:cond_numerics}
\end{table}

We validate these choices by measuring the thermal conductivity in a simple test problem under the conservative and adopted resolution requirements. Specifically, we set up a 1D planar (Cartesian) slab with no self-gravity and with a specified temperature gradient in hydrostatic equilibrium. The initial temperature profile is set by integrating Fourier's law for a pure $\sigma \propto v^{-4}$ cross section with a steady heat flux. The boundary condition is a reflecting thermal wall (that is, particles which exit the computational domain are reflected with a temperature drawn from a flux-weighted Maxwellian at fixed temperature). 

To compare with the theoretical expectation, we integrate Fourier's law for 
\begin{equation}
    \int \frac{1}{\kappa} \dl x = - \int \frac{1}{q} \dl \nu^2
\end{equation}
This formulation avoids noisy numerical derivatives, and does not require an assumed temperature profile. After an equilibration period of two thermal box crossing timescales, we measure the heat flux $q$ as in \cref{sec:fluid}. Then, we compute 
\begin{equation}
   \mathcal{R}_{\mathrm{CE,2}} \equiv -\frac{\bar{q}}{\Delta \nu^2}{\int\frac{\dl x}{\kappa_{\mathrm{CE},2}}}, 
\end{equation}
where $\Delta T$ is the temperature variation across the interior of the box (that is, omitting an 8-cell Knudsen layer at each boundary). The measurement of $q$ is averaged over the entire box, since in steady state $\difs{q}{x} = 0$.

\begin{table}
    \centering
    \caption{Setup and measurement parameters of the thermal conductivity test, in code units:
    length $\tilde{l}$, velocity $\tilde{v}$ and time $\tilde{t}=\tilde{l}/\tilde{v}$.}
    \smallskip
    \begin{tabular}{ll}
        \toprule
        Quantity & Value\\
        \midrule
        Box length $L$ & $0.2\,\tilde{l}$ \\
        Wall temperatures $\nu_{\mathrm{lo}}^2$, $\nu_{\mathrm{hi}}^2$ & $1$, $1.5\,\tilde{v}^2$ \\
        Cross section $\sigma_{\mathrm{m}}(v)$ & $\sigma_0/[1+(v/w)^2]^2$ \\

        Column density per particle & $50\,\tilde{l}^{-2}$ \\
        Number of particles & \num{3.1e5} \\
        \bottomrule
    \end{tabular}
    \label{tab:cond_setup}
\end{table}

\begin{figure}
    \centering
    \includegraphics[width=\linewidth]{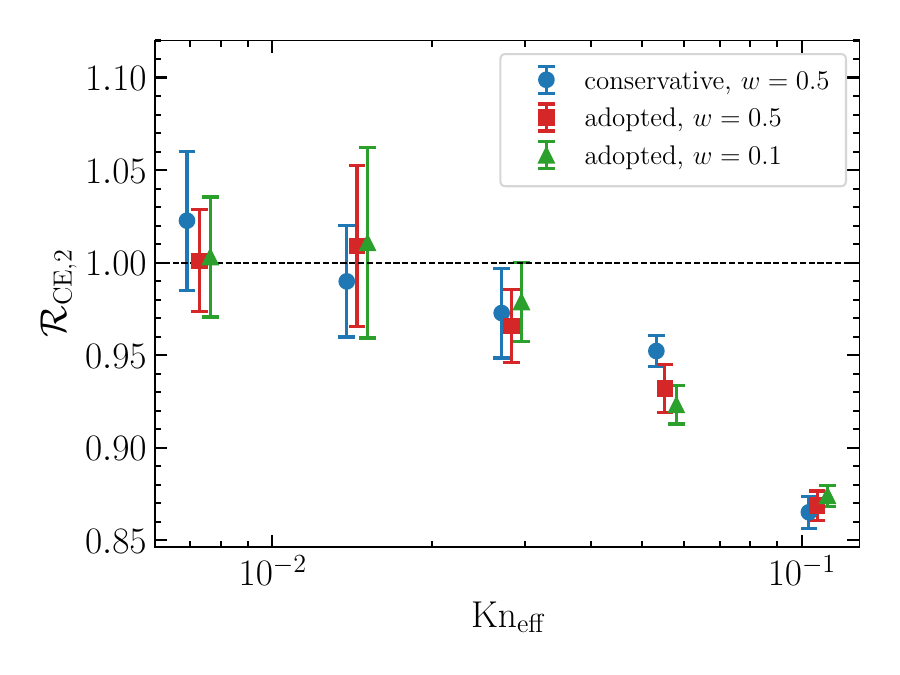}
    \caption{The integral ratio $\mathcal{R}_{\mathrm{CE,2}}$ for the adopted numerics based on $\tilde{\sigma}_{\mathrm{m}}$ and the total collision rate (with $w=0.5$ and $w=0.1$) and for the conservative numerics (with $w=0.5)$, for several Knudsen numbers. The points are offset slightly horizontally to improve legibility.}
    \label{fig:conduct_meas}
\end{figure}

We perform this measurement using the adopted and conservative prescriptions  for five cross sections from the deep SMFP to transition range (here the effective Knudsen number is $\mathrm{Kn}_{\mathrm{eff}}\equiv \lambda_{\mathrm{eff}}/L$). The box parameters are described in \cref{tab:cond_setup}. We estimate the uncertainty of this measurement by the error on the mean, using the variance calculated on $\mathcal{R}_{\mathrm{CE,2}}$ using contiguous blocks of roughly twenty snapshots (which are much larger temporally than either the collision or the cell-diffusion timescales). Except for the deepest SMFP test, these simulations use a total of \num{180} snapshots each separated by 10 thermal cell-crossing times. However, as $\mathrm{Kn}$ shrinks the heat flux decreases and the measurement becomes increasingly noisy. Thus, for the lowest-$\mathrm{Kn}$ simulations we perform the same measurement but from seven repetitions of the simulation (for a total of \num{1260} snapshots). The conservative runs can be completed at reasonable computational cost only for $w=0.5$. Using the adopted numerics, we have simulated both $w=0.5$ and $w=0.1$. 

\Cref{fig:conduct_meas} shows that adopting the \enquote{effective} rather than conservative resolution requirements does not significantly bias the thermal conductivity. As expected, as the mean free path grows, the conductivity drops below the Chapman–Enskog value. The robustness of these numerical prescriptions is additionally supported (albeit circumstantially) by the close agreement in the deep SMFP between the velocity-dependent fluid simulations and the {\codename} simulations: in that regime the free parameters of the fluid model have a minor impact on the collapse time. 

Recently, \citet{Ramos:2025lvk} showed that coarse-graining the distribution function gives rise to a \enquote{macroscopic} transport cross section which differs from the \enquote{microscopic} one. However, the coarse-graining procedure defined in that work differs conceptually from the stochastic sampling of the distribution function typical of $N$-body methods. This conductivity test shows empirically that the DSMC method recovers the correct transport properties from the microscopic cross section (which is a standard result in computational fluid dynamics 
\cite{bird1994molecular}).

\clearpage

\end{document}